\documentclass{emulateapj}

\usepackage{enumerate}
\usepackage{ulem,color}

\shorttitle{Measuring the Alfv\'enic Nature of the Interstellar Medium}
\shortauthors{Burkhart et al.}

\begin{document}

\title{Measuring the Alfv\'enic Nature of the Interstellar Medium: Velocity Anisotropy Revisited}

\author{Blakesley Burkhart\altaffilmark{1}, A. Lazarian\altaffilmark{1},  I. C. Le\~ao\altaffilmark{2}, J. R. de Medeiros\altaffilmark{2}, \&  A. Esquivel\altaffilmark{3} }
\affil{$^1$ {Astronomy Department, University of Wisconsin-Madison, 475 North Charter Street, Madison, WI 53706-1582, USA}}
\affil{$^2$ {Departamento de F\'isica, Universidade Federal do RN 59072-970 Natal-RN, Brazil}}
\affil{$^3$Instituto de Ciencias Nucleares, Universidad Nacional Autonoma de Mexico, Apartado Postal 70-543,04510 Mexico D.F., Mexico}

\begin{abstract}
The dynamics of the interstellar medium (ISM) are strongly affected by turbulence, which shows increased anisotropy in the presence of a magnetic field. 
We expand upon the Esquivel \& Lazarian method to estimate the Alfv\'en Mach number using the structure function anisotropy in velocity centroid data from position-position-velocity maps. 
We utilize 3D magnetohydrodynamic (MHD) simulations of fully developed turbulence, with a large range of sonic and Alfv\'enic Mach numbers, 
 to produce synthetic observations of velocity centroids with observational characteristics such as thermal broadening, cloud boundaries,
noise, and radiative transfer effects of carbon monoxide. In addition, we investigate how the resulting anisotropy-Alfv\'en Mach number dependency found in 
Esquivel \& Lazarian (2011) might change
when taking the second moment of the position-position-velocity cube or when using different expressions to calculate the velocity centroids.
We find that the degree of anisotropy is related primarily to the magnetic field strength (i.e. Alfv\'en Mach number) and
 the line-of-sight orientation, with a secondary effect on sonic Mach number. If the line-of-sight is parallel to up to $\approx$45 deg off of the mean field direction, the velocity centroid anisotropy
is not prominent enough to distinguish different Alfv\'enic regimes. The observed anisotropy is not strongly affected by including radiative transfer, although future studies
should include additional tests for opacity effects.  These results open up the possibility of studying the magnetic nature of the ISM using statistical methods
in addition to existing observational techniques. 
\end{abstract}


\section{Introduction}

The current picture of the interstellar medium (ISM) vitally includes magnetohydrodynamic (MHD) turbulence acting on scales ranging from kiloparsecs to sub-AU
(see Armstrong et al. 1995; Elmegreen \& Scalo 2004).
This is in part due to the fact that MHD turbulence is of key importance
for fundamental astrophysical processes, e.g. heat transport,  star
formation, Galactic pressure support, magnetic reconnection and the acceleration of cosmic rays.

MHD turbulence is notoriously difficult to study both observationally and theoretically (see Elmegreen \& Scalo 2004 for further discussion).  
In light of this, numerical simulations have
tremendously influenced our understanding of the physical
conditions and statistical properties of MHD turbulence (see
Mac Low \& Klessen 2004, Ballesteros-Paredes et al. 2007,
McKee \& Ostriker 2007 and ref. therein). Present codes can
produce simulations that resemble observations in terms of structures and scaling laws, but because
of their limited numerical resolution, they cannot reach the observed Reynolds\footnote{The Reynolds number is $Re\equiv L_fV/\nu=(V/L_f)/(\nu/L^2_f)$ which is the ratio of an eddy 
turnover rate $\tau^{-1}_{eddy}=V/L_f$ and the viscous dissipation rate $\tau_{dis}^{-1}=\eta/L^2_f$. Therefore large $Re$ correspond to 
negligible viscous dissipation of large eddies over the cascading time $\tau_{casc}$ which is equal to $\tau_{eddy}$ in Kolmogorov turbulence.} numbers of the ISM.

Statistical studies represent the best hope to bridge the gap between simulations and observations.
Thus, many techniques beyond the traditional turbulence power spectrum have been developed to study and parameterize observational magnetic
turbulence. These include higher order spectra, such as
the bispectrum (Burkhart et al. 2009), higher order statistical moments (Kowal, Lazarian, \&
Beresnyak 2007; Burkhart et al. 2010), topological techniques (such as genus, see Chepurnov \& Lazarian 2009), clump and hierarchical structure algorithms (such as dendrograms, see Goodman et al. 2009;
Burkhart et al. 2013a), principle component analysis (i.e. PCA, Heyer \& Schloerb 1997; Heyer et al. 2008), Tsallis function studies for ISM turbulence (Esquivel \& Lazarian 2010; Tofflemire, Burkhart, \& Lazarian 2011), 
Velocity Channel Analysis and Velocity Coordinate Spectrum (Lazarian \& Pogosyan 2004, 2006, 2008),
and structure/correlation functions as tests of intermittency and anisotropy (Cho \& Lazarian 2003; Esquivel \& Lazarian 2005; Kowal \& Lazarian 2010).

However, the results of these statistical applications to numerics and/or observations are less insightful without
being placed into the theoretical framework of turbulence.
The famous Kolmogorov (1941) theory of turbulence describes the hydrodynamic counterpart of MHD turbulence.
The transfer of energy from large-scale eddies to smaller scales continues without losses until the cascade reaches eddies that are small enough to dissipate energy over an eddy turnover time. 
In terms of the ISM, the injection scale and main energy sources are still unknown, but it is clear that
turbulence in the Galaxy is driven on large scales (kiloparsec) by supernova, galactic fountain, high-velocity cloud impacts, hydrodynamical and magnetohydrodynamical instabilities, or some combination of these.  
At small scales one should see the scales corresponding to sinks of energy, i.e. dissipation of energy. 

The ISM is also magnetized, and therefore Alfv\'enic perturbations are vital to the development of an MHD cascade. 
Contrary to Kolmogorov turbulence, in the presence of a dynamically important magnetic field, turbulent eddies become elongated along the mean magnetic field (i.e. they become anisotropic)
and Alfv\'enic perturbations develop an independent cascade which
proceeds perpendicular to the local magnetic field and is marginally affected
by the fluid compressibility (see Cho \& Lazarian 2003).
The dynamic influence of the magnetic field and the induced anisotropy of the eddies increases as the cascade proceeds down to smaller scales.
 This corresponds to the predictions of the Goldreich \& Sridhar (1995, henceforth GS95) theory of Alfv\'enic turbulence.
 
It is important to stress that the above picture of magnetized turbulence is developed in the context of the \textit{local magnetic field} relative to the eddies. The anisotropy
of interstellar turbulence that is accessible to observations, which are averaged along the line-of-sight (LOS), is sampled in the global reference frame, relative
to the large scale mean magnetic field. In this case, the anisotropy is determined by the largest scale eddies. 
The first discussion of the possibilities of observationally measuring the large scale anisotropy induced by the magnetic field
was  in Lazarian, Pogosyan \& Esquivel (2002, henceforth LPE02), who proposed to measure contours of equal correlation corresponding to data within different velocity
channel thickness in HI data (i.e. in a similar procedure to the Velocity Channel Analysis).
Follow up papers by Esquivel \& Lazarian (2005) and Esquivel \& Lazarian (2011) showed that velocity centroids can be used
for testing whether turbulence is sub- or super-Alfv\'enic while  Heyer et al. (2008) used PCA to 
recover a empirical relationship for the anisotropy found in simulations and applied this to molecular cloud observations.

Esquivel \& Lazarian (2011, henceforth EL11) developed a method to quantify the large scale anisotropy of the turbulent cascade in velocity centroid maps of Position-Position-Velocity 
(PPV) data cubes.  In this paper, we expand upon the EL11 method in terms of the parameter space studied and the applicability of their method to the observations.
EL11 showed that this method is highly sensitive to the global Alfv\'enic Mach number of turbulence, defined as $M_A \equiv \langle V_L / v_A \rangle$, as well as the direction of the magnetic field.
We investigate a similar approach, but now use higher resolution simulations and consider more realistic synthetic observations.
Furthermore, we study the relation of the isotropy degree with the Alfv\'en Mach number ($M_A$) and the sonic Mach number ($M_s=V_L /c_s$, where $c_s$ is the sound speed) for different LOS orientation angles, which can be useful for estimating the 3D structure of the magnetic field in the ISM.
The paper is organized as follows.  In \S~\ref{sec:theory} we describe the GS95 model for sub- and super-Alfv\'enic turbulence and how this applies to the EL11 method,
In \S~\ref{secmhd} we describe our extended data base of MHD simulations and describe in detail our procedure to create synthetic observations and calculate the isotropy degree. 
In \S~\ref{secres} we describe our results followed by the discussion in \S~\ref{secdisc} and conclusions in \S~\ref{seccon}.

\section{The GS95 Anisotropy for sub-Alfv\'enic and super-Alfv\'enic Turbulence}
\label{sec:theory}
The GS95 theory assumes the  injection of energy at scale $L$ and the injection velocity equal to the Alfv\'en velocity in
the fluid $V_A$, i.e. the Alfv\'en Mach number $M_A\equiv (V_L/V_A)=1$ (i.e. trans-Alfv\'enic turbulence) , where $V_L$ is the injection velocity.
The GS95 model was later generalized
for both sub-Alfv\'enic, i.e. $M_A<1$, and super-Alfv\'enic, i.e. $M_A>1$, cases (see Lazarian \& Vishniac 1999 and Lazarian 2006)
and thus the results of EL11 and the current work must be understood in this context.

For the eddies perpendicular to the magnetic field, the original Kolmogorov energy scaling is applicable resulting in perpendicular motions 
scaling as $V_l\sim \l_{\bot}^{1/3}$, where $l_{\bot}$ denotes eddy scales measured perpendicular to the local
magnetic field. 
Mixing motions induce Alfv\'enic perturbations that determine the parallel size of the magnetized eddy.  
This concept of {\it critical balance}  i.e. the equality of the eddy turnover time ($l_{\bot}/v_l$) and the period of 
the corresponding Alfv\'en wave $\sim l_{\|}/V_A$, where $l_{\|}$ is the parallel eddy scale and $V_A$ is the Alfv\'en velocity. 
Making use of $v_l\sim \l_{\bot}^{1/3}$, one finds the scaling relation for the parallel and 
perpendicular eddies as: $l_{\|}\sim l_{\bot}^{2/3}$. 
This reflects the scale-dependent anisotropy of eddies along the magnetic field lines
as the energy cascades proceeds to smaller scales and has been tested using 2nd order structure functions in the reference frame to the 
local magnetic field (see Cho \& Lazarian 2003; Beresnyak, Lazarian \& Cho 2005; Kowal \& Lazarian 2010).

The EL11 method takes advantage of the global anisotropy observed in the largest-scale eddies that can be measured via structure function analysis.
The first mention of the use of the structure function anisotropy technique to study turbulence and the direction of the mean magnetic field was made by LPE02, who used
 synthetic spectral line emission maps obtained via MHD turbulence simulations to demonstrate the method's promise.  
Later studies (e.g. Esquivel et al. 2003; Vestuto et al. 2003; Heyer et al. 2008) confirmed that the  anisotropy is evident
from two-point statistics, i.e. the structure function, of observational quantities such as velocity centroids.
While EL11 studied the relation between the anisotropy and the global Alfv\'enic Mach number, they did so with limited resolution simulations and less attention to observational effects such
as thermal broadening and unknown LOS angle relative to the mean magnetic field.  We expand the parameter range of their study and include additional measures and observational considerations.



The EL11 method takes advantage of the global anisotropy observed in the largest scale eddies (which occur at $l_A$, L, or $l_{trans}$ depending on whether turbulence is super, trans, or sub-Alf\'venic, respectively).
The first mention of the use of structure function anisotropy to study turbulence and the direction of the mean magnetic field was made by LPE02, who used
 synthetic spectral line emission maps obtained via MHD turbulence simulations to demonstrate the method's promise.  
Later studies (e.g. Esquivel et al. 2003; Vestuto et al. 2003; Heyer et al. 2008) confirmed that the  anisotropy is evident
from two-point statistics, i.e. the structure function, of observational quantities such as velocity centroids.
While EL11 studied the relation between the anisotropy and the global Alfv\'enic Mach number, they did so with limited resolution simulations and less attention to observational effects such
as thermal broadening and unknown LOS angle relative to the mean magnetic field.  We expand the parameter range of their study and include additional measures and observational considerations.

\section{MHD simulations and structure functions of synthetic velocity centroid maps}
\label{secmhd}

We generate  3D numerical simulations of isothermal compressible (MHD)
turbulence by using the Cho \& Lazarian  (2003) MHD code and varying the input
values for the sonic and Alfv\'enic Mach number.  Turbulence is driven with large-scale
solenoidal forcing.  The magnetic field has contributions from a uniform background field and a
fluctuating turbulent field: ${\bf B}= {\bf B}_\mathrm{0} + {\bf b}$. Initially ${\bf b}=0$.
The simulations have resolutions of either 512$^3$ or 256$^3$ and the models are run
for $t \sim 5$ crossing times, to guarantee full development of the energy cascade.
For more details see Cho \& Lazarian (2003), Kowal, Lazarian, Beresnyak (2007) and Burkhart et al. (2009) and EL11.

We partition our models into three groups corresponding to their Alfv\'enic Mach number, which covers 
sub-Alfv\'enic ($B_\mathrm{0}=5.0, 3.0$) to trans-Alfv\'enic ($B_\mathrm{0}=1.0$) to
super-Alfv\'enic ($B_\mathrm{0}=0.1$) turbulence.
The initial conditions were defined with $\rho = 1$ and the Alfv\'en speed $v_a = |\mathbf{B}| / \sqrt{4\pi\rho}$.
The simulations were evolved to reach a stationary state with the rms velocity close to unity ($v_{\rm rms} \sim 0.7$).
For each group we compute several models with different values of
the sonic Mach number (see Table \ref{tabmodels}, second column).
The models are listed and described in Table~\ref{tabmodels}, where $\langle P_{{\rm gas},0} \rangle$ and $B_{0}$ represent
the initial gas pressure and magnetic field, respectively. 
The labels given by EL11 for 6 models (M1--M3 and M7--M9) with similar initial conditions as here are indicated in parentheses.
We note that the units on these quantities are given in dimensionless code units.  For a detailed discussion on how to convert
code units into physical units, see Appendix A of Hill et al. (2008).

\begin{table} 
  \scriptsize
  \centering
    \begin{tabular}{l|c|c|c|c|c}
     \hline\hline
      Model  &  $B_{0}$  &  $\langle{P_{{\rm gas},0}}\rangle$  &  $M_s$  &  $M_A$  & Resolution \\
     \hline
      L1      &  0.1  &  0.0049 &  $\sim$8.7  &  $\sim$6.1  &  512$^3$ \\
      L2(M1)  &  0.1  &  0.01   &  $\sim$5.7  &  $\sim$5.7  &  512$^3$ \\
      L3      &  0.1  &  0.025  &  $\sim$3.5  &  $\sim$5.5  &  512$^3$ \\
      L4      &  0.1  &  0.05   &  $\sim$2.5  &  $\sim$5.6  &  512$^3$ \\
      L5(M2)  &  0.1  &  0.1    &  $\sim$1.8  &  $\sim$5.8  &  512$^3$ \\
      L6(M3)  &  0.1  &  1.0    &  $\sim$0.6  &  $\sim$5.8  &  512$^3$ \\
      L7      &  0.1  &  2.0    &  $\sim$0.4  &  $\sim$5.4  &  512$^3$ \\
     \hline
      L8      &  1.0  &  0.0049 &  $\sim$8.0  &  $\sim$0.6  &  512$^3$ \\
      L9     &  1.0  &  0.0077 &  $\sim$6.3  &  $\sim$0.6  &  512$^3$ \\
      L10(M7) &  1.0  &  0.01   &  $\sim$5.5  &  $\sim$0.5  &  512$^3$ \\
      L11     &  1.0  &  0.025  &  $\sim$3.5  &  $\sim$0.6  &  512$^3$ \\
      L12     &  1.0  &  0.05   &  $\sim$2.6  &  $\sim$0.6  &  512$^3$ \\
      L13(M8) &  1.0  &  0.1    &  $\sim$1.8  &  $\sim$0.6  &  512$^3$ \\
      L14(M9) &  1.0  &  1.0    &  $\sim$0.6  &  $\sim$0.6  &  512$^3$ \\
      L15     &  1.0  &  2.0    &  $\sim$0.4  &  $\sim$0.6  &  512$^3$ \\
     \hline
      L16     &  3.0  &  0.01   &  $\sim$10.2 &  $\sim$0.3  &  256$^3$ \\
      L17     &  3.0  &  0.1    &  $\sim$3.1  &  $\sim$0.3  &  256$^3$ \\
      L18     &  3.0  &  1.0    &  $\sim$1.0  &  $\sim$0.3  &  256$^3$ \\
     \hline
      L19     &  5.0  &  0.01   &  $\sim$8.7  &  $\sim$0.2  &  256$^3$ \\
      L20     &  5.0  &  0.1    &  $\sim$2.6  &  $\sim$0.2  &  256$^3$ \\
      L21     &  5.0  &  1.0    &  $\sim$0.8  &  $\sim$0.2  &  256$^3$ \\
     \hline
    \end{tabular}
    \caption{Parameters of the MHD Simulations. Note: labels in parentheses are those given by EL11 for simulations with similar initial conditions.}
  \label{tabmodels}
\end{table}

We create synthetic PPV cubes and the corresponding velocity centroid maps assuming a fully
optically thin media with the emissivity being proportional to the density.
EL11 computed the velocity centroid maps directly from the simulated cubes as:
\begin{equation}
  C_z(x,y) \equiv \int{\rho(x,y,z) V_z(x,y,z) dz} / \int{\rho(x,y,z) dz} .
\label{eq:ideal_centroids}
\end{equation}
This form, henceforth known as the ``ideal centroid'', is convenient to be applied directly to the simulation data described in Sect.~\ref{secmhd}, 
but not on observed PPV cubes.
In this work, we consider the velocity centroid 
which is the definition applied on observed PPV cubes.
We first computed a synthetic PPV cube from the intensity distribution $I(\mathbf{X}) \equiv \int{\rho_s dV_{LOS}}$,
where $\rho_s$ is the density of emitters in the PPV space (i.e. the intensity values of the 3D PPV data cube),
{\bf X} denotes the position on the plane of the sky, and $V_{LOS}$ is the LOS velocity-axis. 
The integral is made in velocity along the entire LOS, applied at every position in the plane of the sky. For instance, in the expression in equation (\ref{eq:ideal_centroids}), the LOS coincides with the $z$-axis, and $\mathbf{X}=(x,y)$.


\begin{figure*}
   \centering

   \includegraphics[scale=0.47]{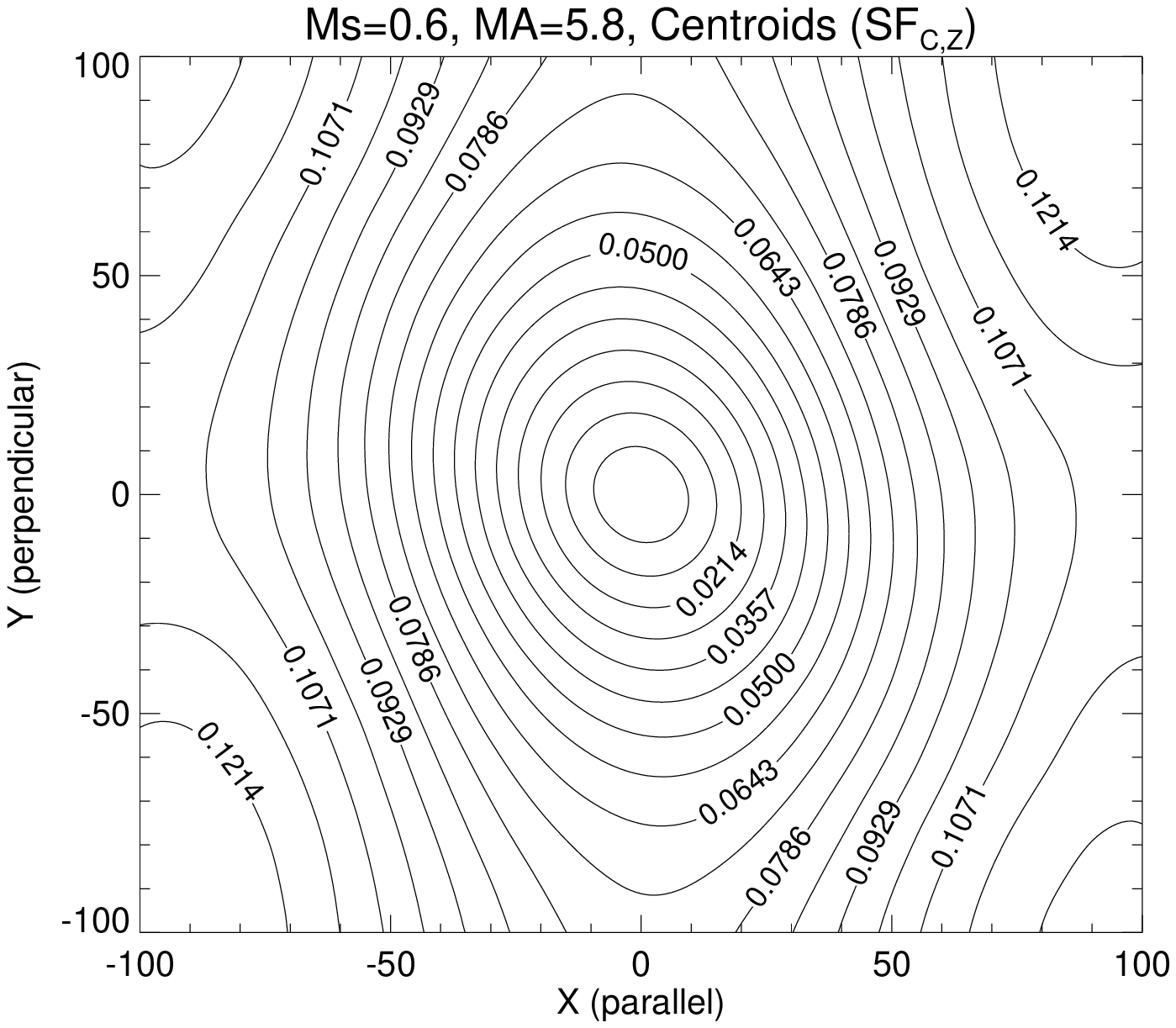}
   \includegraphics[scale=0.47]{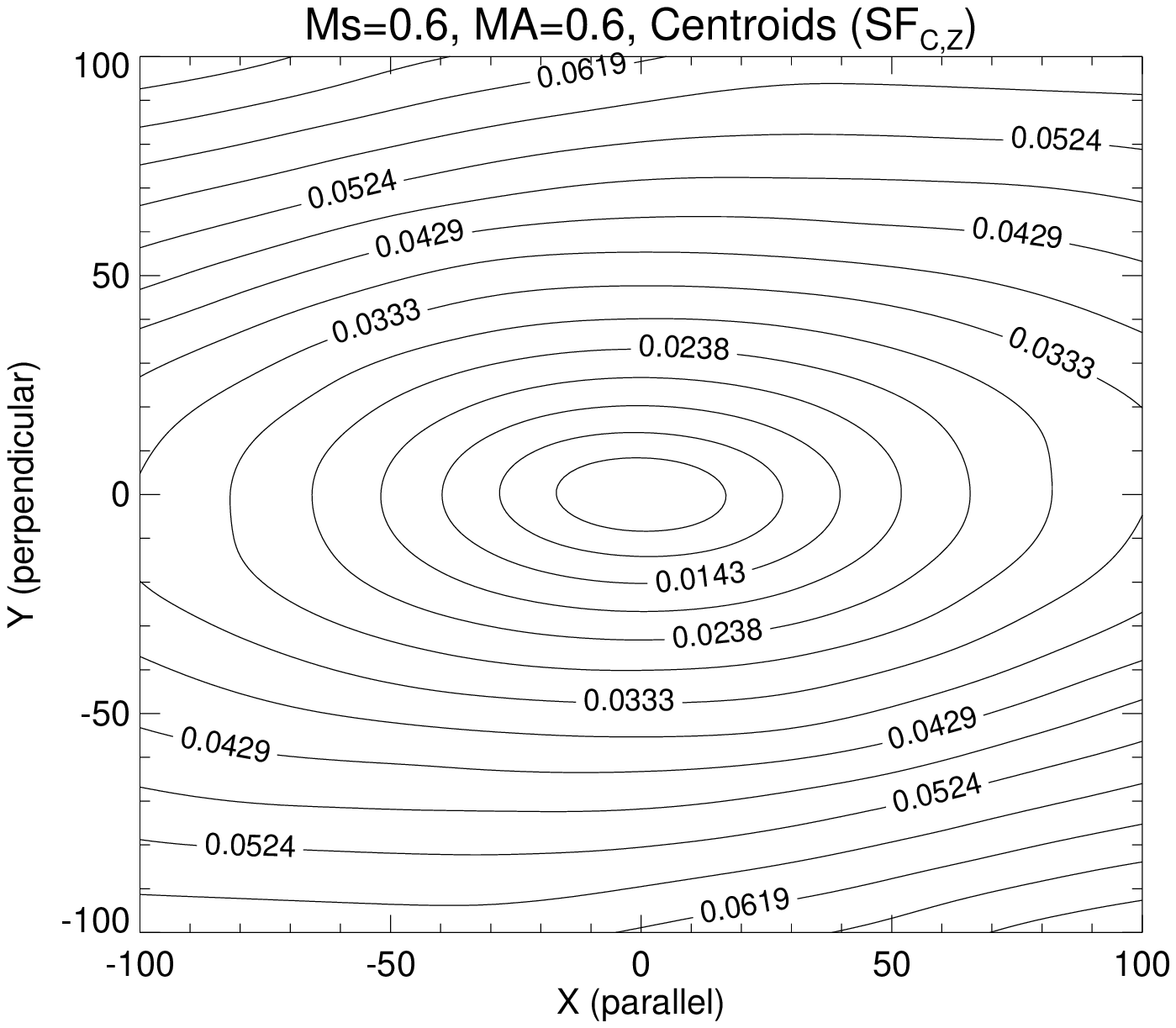}
\caption{Example of the isocontours of the structure function for simulations L7 (left) and L15 (right).
For both simulations, the LOS is perpendicular to the mean B-field, which is, in this case, aligned with the horizontal X-axis.
Isotropic eddies can be found in L7, which has a low value of magnetic field while eddies become more anisotropic in L15, which has a
higher value of magnetic field.}
   \label{fig1}
   \end{figure*}

We compute the structure function as $SF(\mathbf{r}) = \langle [f(\mathbf{x}) - f(\mathbf{x} + \mathbf{r})]^2 \rangle$.
We  denote  the structure function  of velocity centroids obtained from the PPV cube with LOS along
the $x$-axis (which in our simulations is parallel to the mean magnetic field) as $SF_{C,x}(\mathbf{R})$. 
$SF_{C,y}(\mathbf{R})$ and $SF_{C,z}(\mathbf{R})$, denote the structure functions of the velocity centroids of PPV data with LOS along $y$ and $z$ axes, respectively (both of
which are perpendicular to the mean magnetic field).
In our application, the structure function is a two-dimensional function, with isocontours that are approximately circular for isotropic Kolmogorov-type turbulence and elliptical for anisotropic turbulence.
We present an example case in Figure \ref{fig1} comparing models L7 and L15 from Table 1. These two simulations have the same sonic Mach number
but almost an order of magnitude difference in Alfv\'en Mach number.  Model L7 (with $M_A=5.8$) has an isotropic (circular) structure function  while model L15 (with
$M_A=0.6$) has and anisotropic (elliptical) structure function. 
As was discussed in Section \ref{sec:theory}, the anisotropy indicates the presence of a magnetic field and increases with increasing magnetic field.
In order to quantify this effect observed in the 2D structure function of the velocity centroid maps,
we define the isotropy degree as being the ratio of the structure functions in two perpendicular directions to the LOS, 
intersecting at the distribution center:
\begin{equation} SF_{C,z}(x,0)/SF_{C,z}(0,y) \end{equation}

In what follows, we will explore the relation of the isotropy degree with $M_A$ and $M_s$ and compare our results with EL11.

\section{Results}
\label{secres}

\begin{figure*}
   \centering
   \includegraphics[scale=.7]{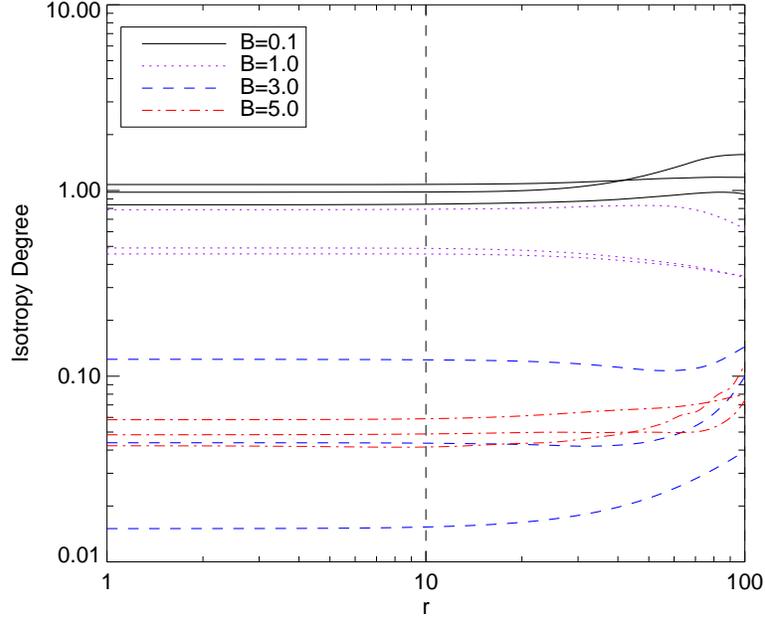}
\caption{Isotropy degree of the velocity centroid structure function vs. r for models L1, L2, L3, L9, L11, L12, L17, L18, L19, L20, L21 and L22.
We chose these models to represent three different sonic Mach number groups from each of the four different Alfv\'enic Mach numbers in our simulation set. }
   \label{figSF2}
   \end{figure*}

\begin{figure*}
   \centering
 \includegraphics[scale=0.7]{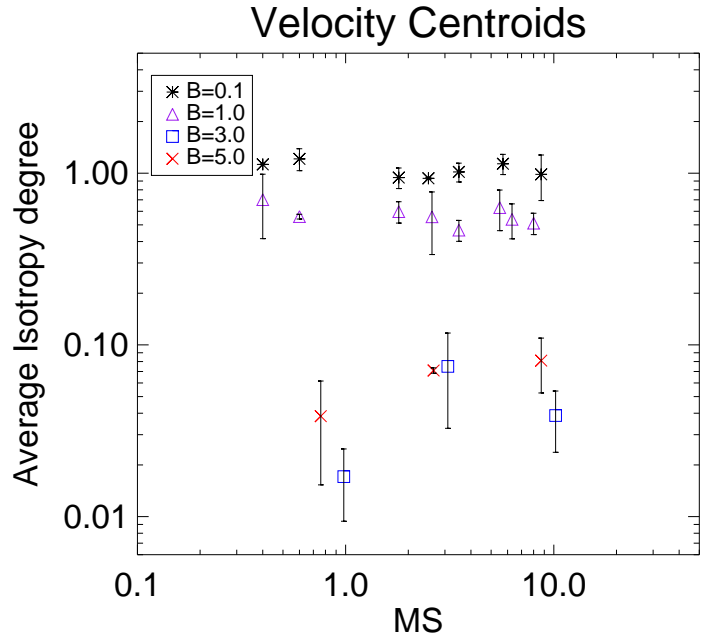}
 \caption{Degree of anisotropy in all the models averaged over scales from 10 grid points to 1/5 of the computational box. 
The x-axis corresponds to the sonic Mach number, and the Alfv\'enic Mach number is indicated 
by the various symbols (and colors in the online version) as shown in the label. 
In all panels the results are obtained by averaging the two cases where the LOS is perpendicular to the mean field (i.e. along the y and z directions in our cubes). 
The error bars show the maximum variation of the averaging procedure (including variation across scales).}
   \label{fig4}
   \end{figure*}

In order to illustrate the general trends, we selected twelve simulations (three different sonic Mach number) for the four different values of magnetic field from our simulation parameter space presented in Table 1.
We plot the isotropy degree vs. the spatial separation (r) in Figure~\ref{figSF2} for the different models. 
Vertical lines show our range of r values for obtaining
the average isotropy degree (shown in Figure~\ref{fig4}).  Below a 10 grid point scale, the density from the
MHD simulations is affected by numerical diffusion, and the effect of noise is more pronounced. 
Past 100 grid points, the simulation is dominated by the injection scale of the turbulence.
We find that, in nearly all cases, the anisotropy
is virtually scale-independent from the small scales up
to separations on the order  1/5 of the computational
box (about half the size of the injection scale). We note that scale independence should not exist
in the local frame of reference to the magnetic field.  However, because we only sample the global frame (large-scale eddies) 
we observe no scale dependency in the anisotropy. 
We consider other lines of sight in the next section.

Figure ~\ref{figSF2} shows a clear separation of the isotropy degree for simulations with different value of magnetization across a range of spatial scales.  The simulations generally cluster in isotropy degree around three Alfv\'enic regimes:
high magnetization (B=3.0, 5.0; sub-Alfv\'enic turbulence),
trans-Alfv\'enic turbulence (B=1.0), and super-Alfv\'enic turbulence (B=0.1). Sub-Alfv\'enic simulations with B=3.0 and 5.0 show the lowest
isotropy degrees, implying that the isocontour values of the structure function as applied to maps of the velocity centroids show a large 
degree of anisotropy along the mean magnetic field.

In Figure \ref{fig4} we show the average degree of isotropy as a function of the sonic Mach number for all the models. 
The results are obtained by averaging the two cases where the LOS is perpendicular to the mean field, and over the range of separations from $r=10$ cells to $1/5$  of the computational box.
The error bars show the maximum variation of the averaging procedure (including variation across scales). 
It is clear from Figure~\ref{fig4} that the degree of anisotropy depends
primarily on the Alfv\'enic Mach number and is generally weakly dependent on the sonic Mach number. 
 One can attribute such dependence to the original density field (i.e. arising from
shocks in supersonic turbulence). Several additional statistical and observational methods exist
to find the sonic Mach number (see Burkhart et al. 2010; Burkhart \& Lazarian 2012) and these can be used to break the slight degeneracy observed in Figure~\ref{fig4}.
However, there are additional observational effects that must be considered such as the LOS orientation of 
the magnetic field, the application of cloud boundaries, and the effects of noise, which we will discuss in the next subsection.


\subsection{Application to synthetic observations}

We first repeat the steps EL11 took to make our results more applicable to observations and later add to these steps.
First, we (and EL11) include two different contributions to mimic observational effects:
an $\alpha r^{-2}$ gradient to induce the effect of cloud boundaries,
and white Gaussian noise.
The white noise was produced using
fractional Brownian motion (fBm) structures with a power spectrum index
(Stutzki et al. 1998; Bensch et al. 2001) set to zero.
As done in EL11, the noise was added to the density (the mean density is 1.0)
with a floor value of 0.01.
Thereafter, we apply a Gaussian convolution on the $V_z$ direction for each $(x,y)$ cube positions to mimic the effects of thermal broadening.
This is a new addition to the technique and was not performed in EL11.
The FWHM of the Gaussian is estimated from the velocity dispersion, $\sigma_{\rm thermal} = \sigma_{\rm turb} / M_s$,
where $\sigma_{\rm turb}$ is the turbulent velocity.

Finally, the ``PPV centroid'', $C_z(x,y)$, can be computed from the PPV cube as:

\begin{equation}
C_z(x,y) \equiv \int{V_{z, ppv} \rho_s dV_{z,ppv}} / \int{\rho_s dV_{z,ppv}} .
\label{eq:ppvcent}
\end{equation}
where $V_{z, ppv}$ is the velocity axis along the PPV data cube along the z LOS in the cube.  In practice, Equation \ref{eq:ppvcent} produces output identical to Equation \ref{eq:ideal_centroids},
however Equation \ref{eq:ppvcent} is an observational method for calculating the velocity centroid map while Equation \ref{eq:ideal_centroids} can only be applied to numerical simulations.

We also consider the effects of anisotropy on the maps of mean LOS velocity (which is not an observable) defined as: 
\begin{equation}
V_z(x,y) \equiv 1/N_z \int{V_z(x,y,z) dz},
\label{eq:meanv}
\end{equation}
This is to compare the statistics of the synthetic velocity centroids with the actual average velocity of the turbulence.

In addition to creating the velocity centroid by taking the first moment with respect to $V$, we also investigate the second moment by instead using $V^2$
in Eq.~(\ref{eq:ppvcent}), thus making the isotropy degree more sensitive to velocity  and less sensitive to density. We will refer to this second moment map as ``PPV$^2$ centroid'',
which was not investigated in EL11.

\begin{figure*}
   \centering
   \includegraphics[width=5.2cm, height=4.2cm]{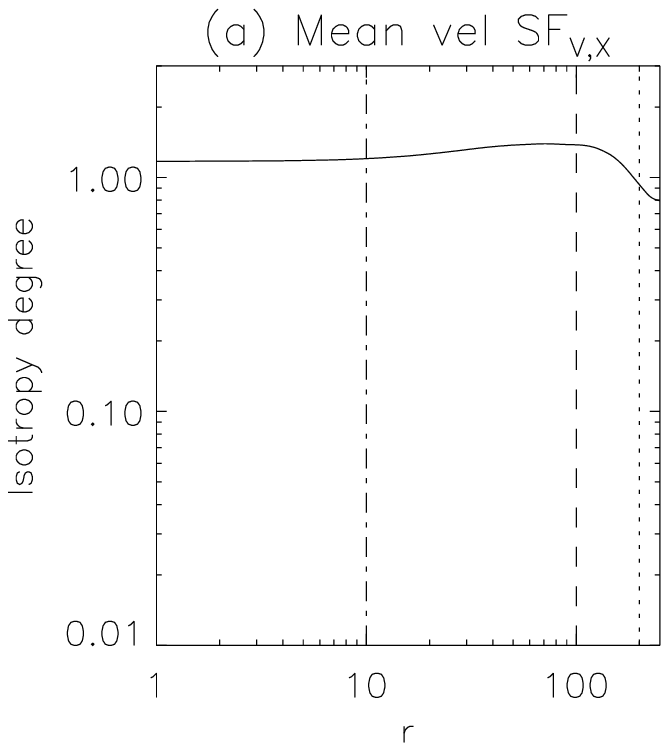}
   \includegraphics[width=5.2cm, height=4.2cm]{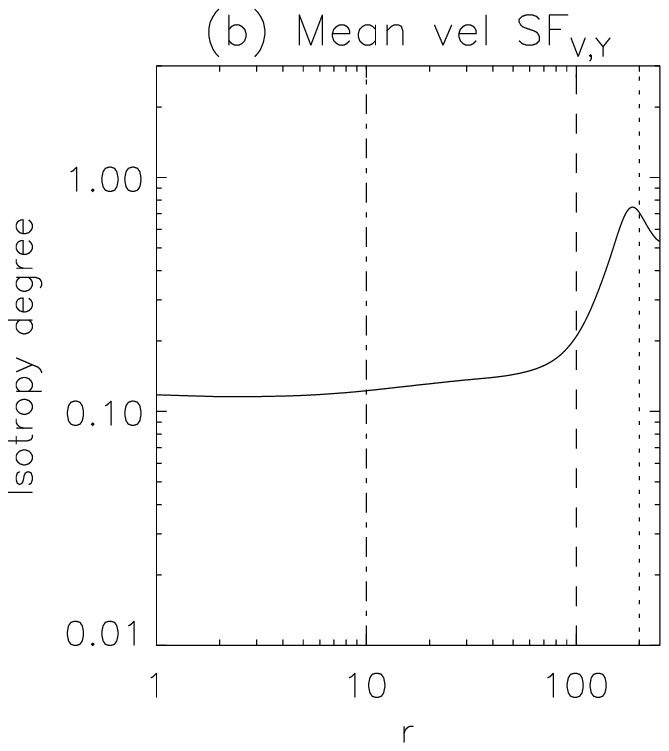}
   \includegraphics[width=5.2cm, height=4.2cm]{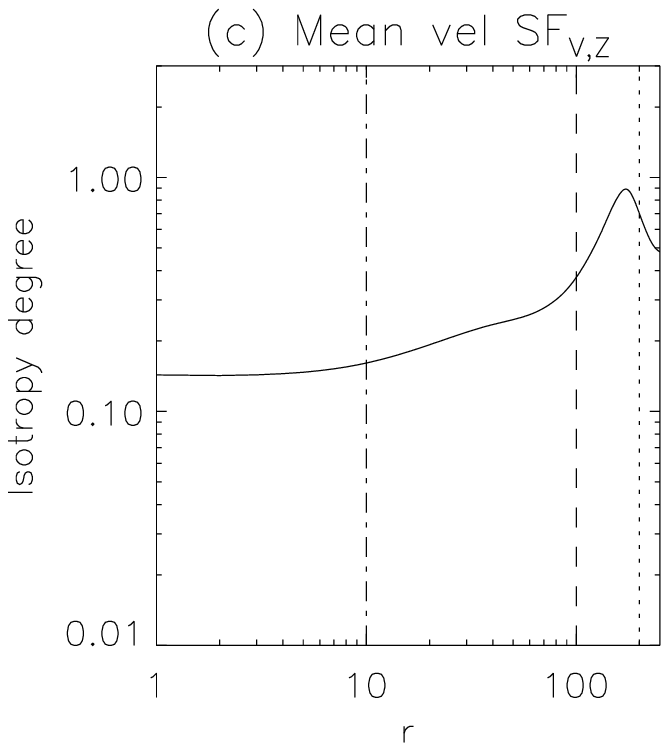}
   \includegraphics[width=5.2cm, height=4.2cm]{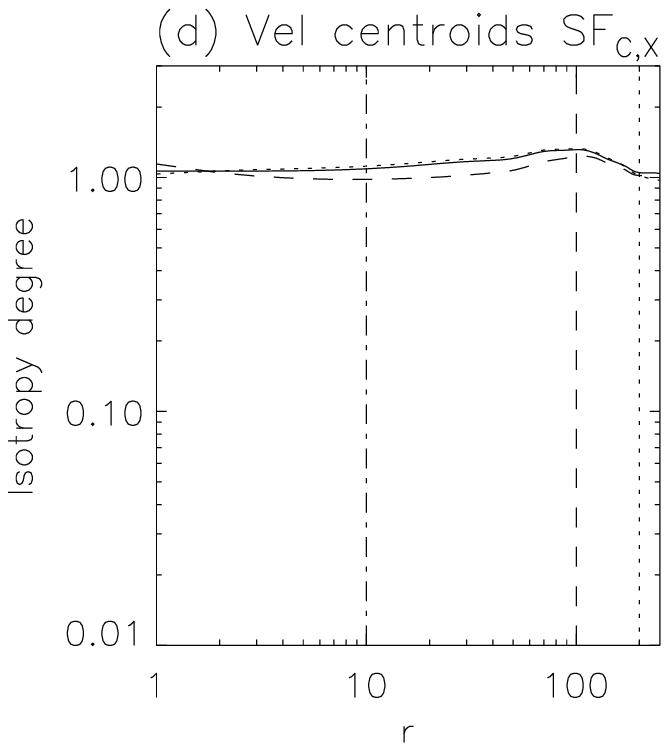}
   \includegraphics[width=5.2cm, height=4.2cm]{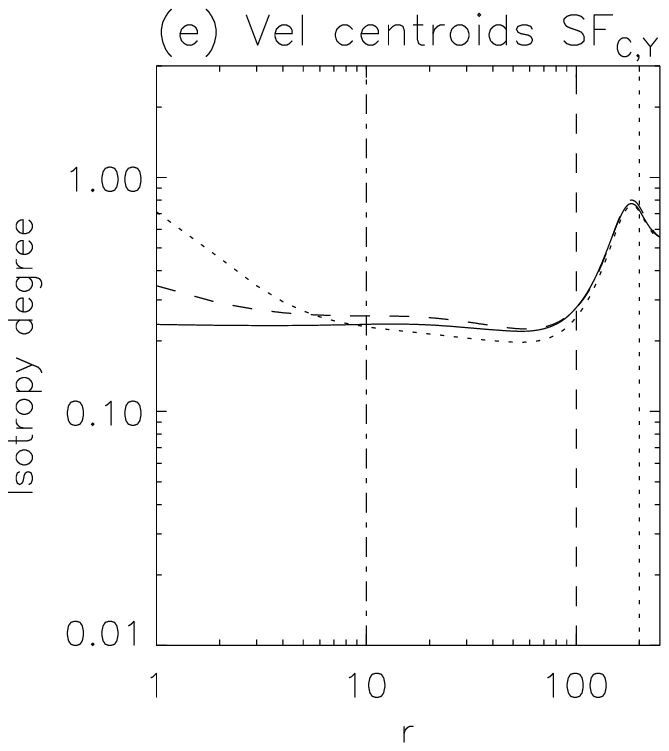}
   \includegraphics[width=5.2cm, height=4.2cm]{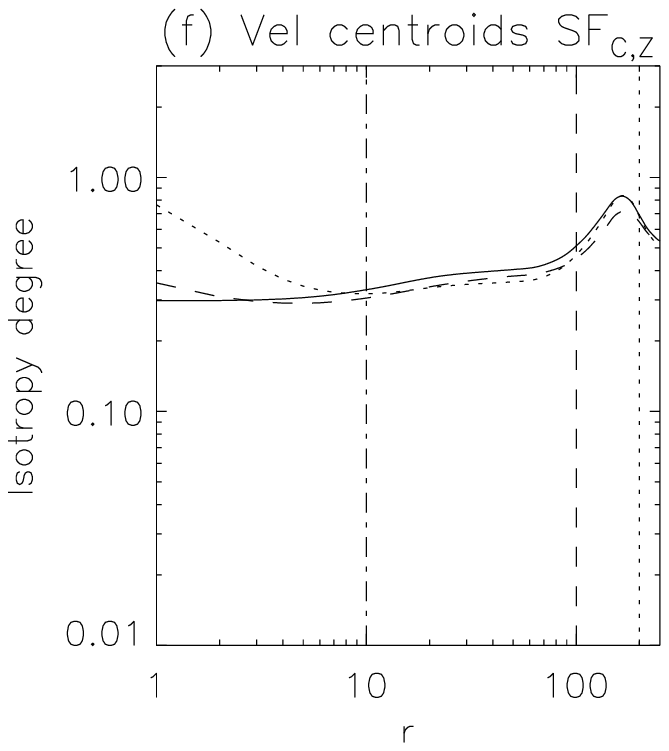}
   \includegraphics[width=5.2cm, height=4.2cm]{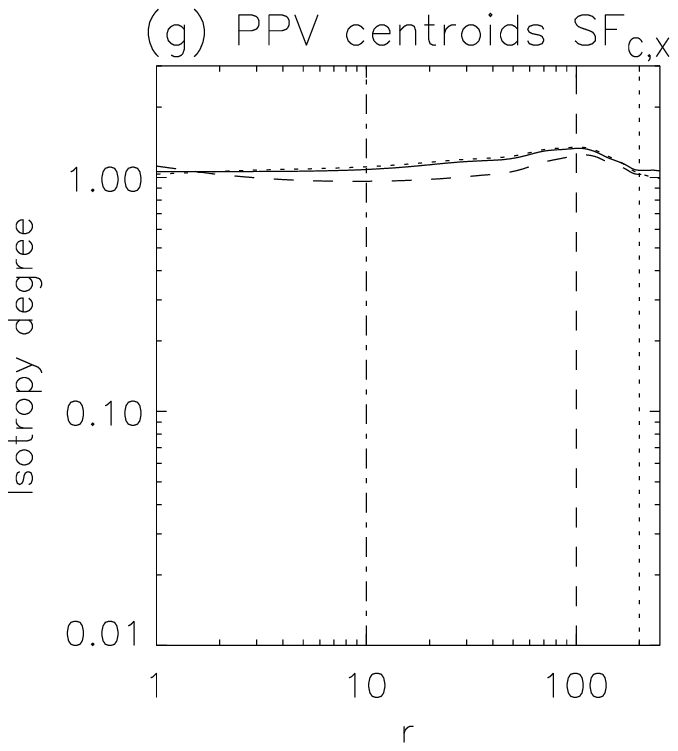}
  \includegraphics[width=5.2cm, height=4.2cm]{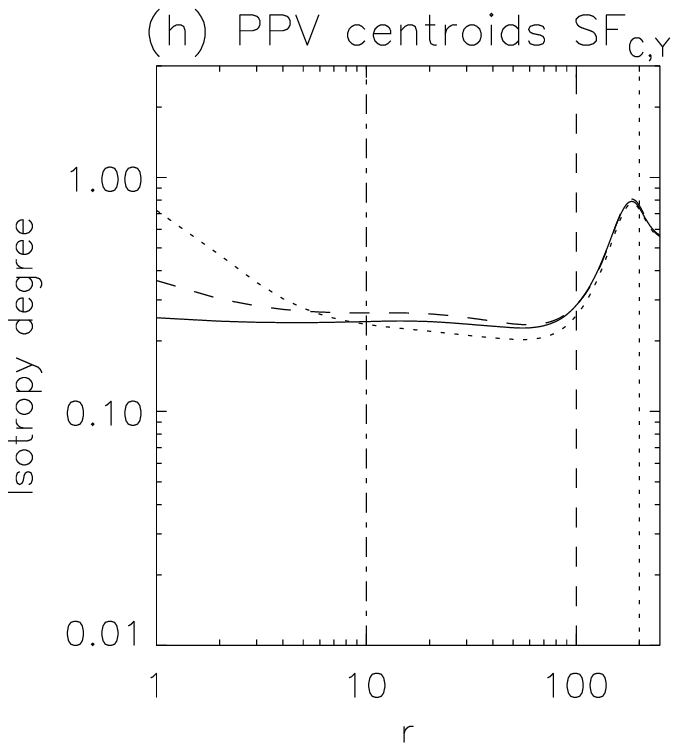}
   \includegraphics[width=5.2cm, height=4.2cm]{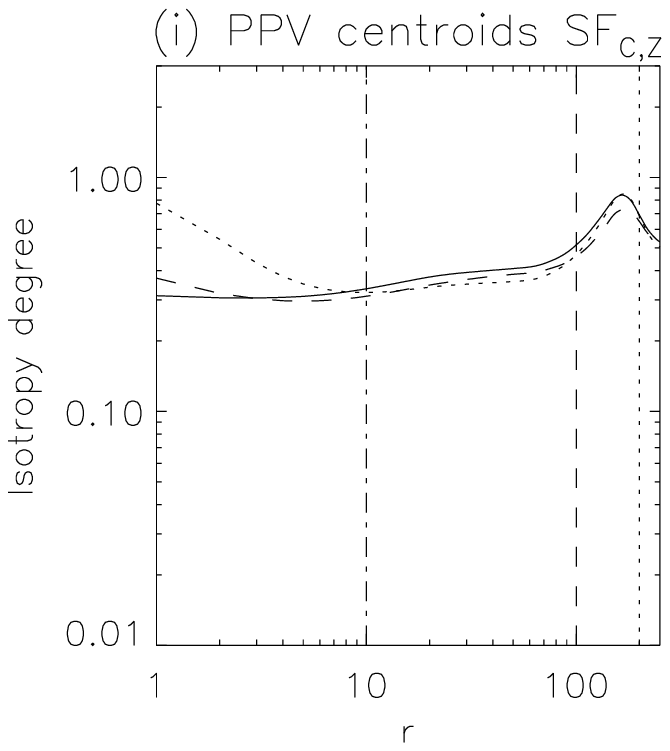}
   \includegraphics[width=5.2cm, height=4.2cm]{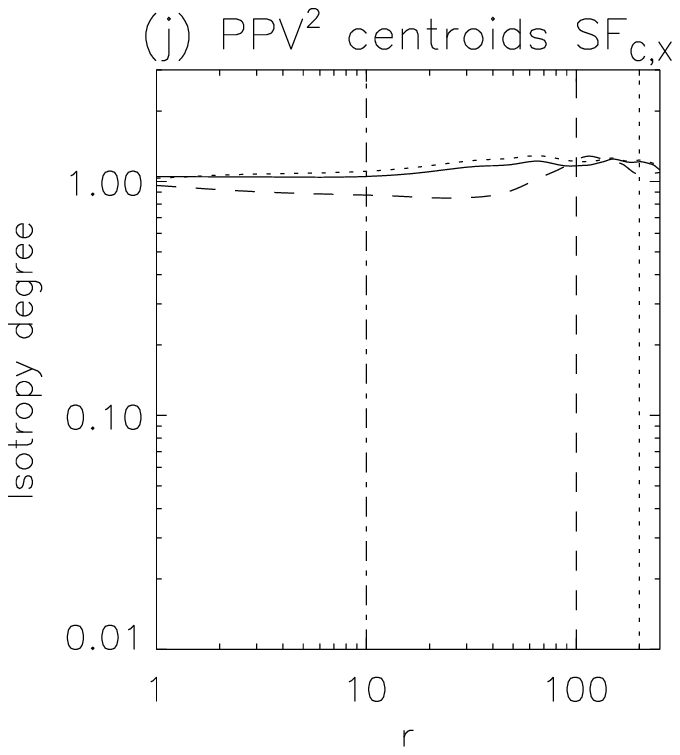}
   \includegraphics[width=5.2cm, height=4.2cm]{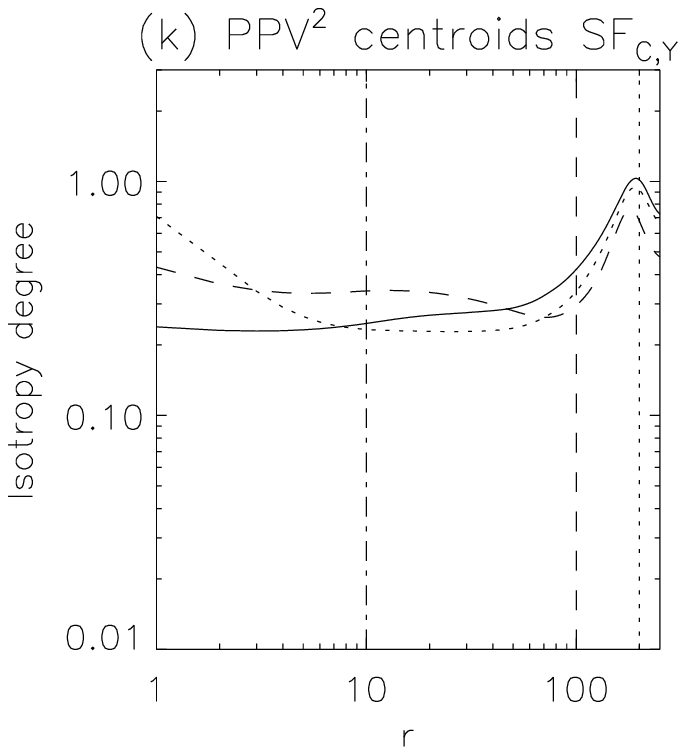}
   \includegraphics[width=5.2cm, height=4.2cm]{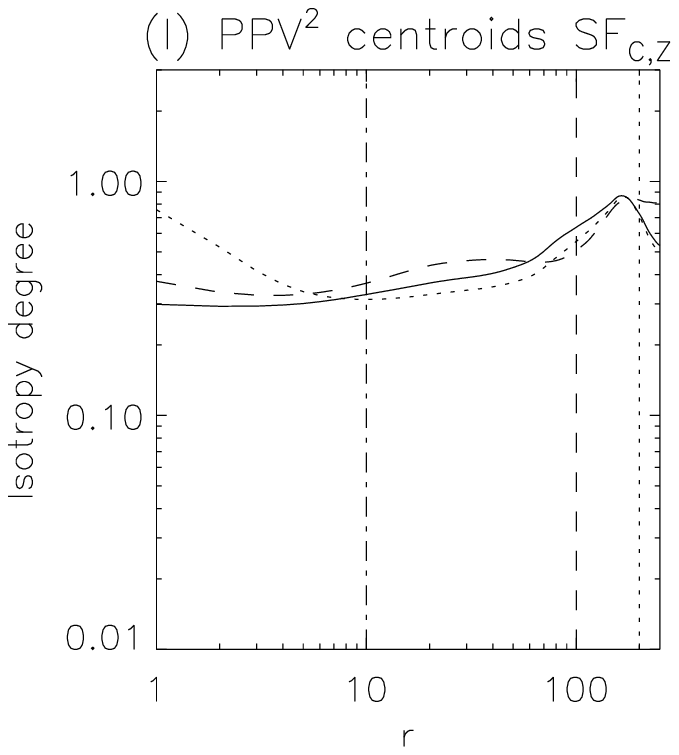}
   \caption{Example of the degree of anisotropy of the structure functions in the same model as those considered by EL11 (ie. model M8), as observed from different directions. The top panels, (a)--(c), were obtained with the mean velocity maps while the middle-upper panels, (d)--(f), were obtained with maps of ``ideal centroids''
 (see Sect.~\ref{secmhd}), all following the same approach as that by EL11. Middle-bottom panels, (g)--(i), were obtained with maps of the ``PPV centroids'' described in Sect. 5.1. 
Bottom panels, (j)--(l), were also obtained from ``PPV centroids'', but by changing the variable $V$ to $V^2$ in Eq.~(\ref{eq:ppvcent}).
The different lines in panels (d)--(i) denote the density field used to obtain the centroids: the solid line corresponds to the original density, the dashed line to the $\alpha r^{-2}$ gradient,
 and the dotted line to the addition of white noise. The LOS is aligned with the x-axis (parallel to the B field) in the left column, panels (a), (d), (g), and (j), with the $y$-axis in the middle column, 
panels (b), (e), (h) and (k), and the $z$-axis in the right column, panels (c), (f), (i) and (l).}
   \label{figSF}
   \end{figure*}

Fig.~\ref{figSF} shows the isotropy degree seen in the structure function  vs. the spatial separation ($r$)
for  a simulation with similar initial conditions as those considered by EL11, i.e. model M8 from Table 1.
We plot different lines of sight relative to the mean magnetic field as columns across.
The top row (panels {\it a, b, c}) shows the structure functions applied to maps of mean velocities $V_x(y,z)$, $V_y(x,z)$ and $V_z(x,y)$.
Panels {\it d, e, f} show the structure functions computed on PPV cubes which are identical to the EL11 method.
Panels {\it g, h, i} show the structure functions applied to the velocity centroids obtained  by taking the first moment of a synthetic PPV cube with thermal smoothing applied
(named ``PPV centroids'').
Finally, in the bottom row (panels {\it j, k, l}), the structure functions were obtained from the second moment denoted PPV$^2$.

The different lines in panels denote the density field ($\rho$) used to obtain the centroids: the solid line corresponds to the original turbulent density field, the dashed line to the density field with an
$\alpha r^{-2}$ gradient applied to it to mimic cloud boundary effects,
and the dotted line to the addition of white Gaussian noise and $\alpha r^{-2}$ cloud boundaries.

The most striking result shown in Figure~\ref{figSF} is that the degree of isotropy is very similar in all rows.
In particular, there is no noticeable difference between panels {\it(d)--(f)} and {\it(g)--(i)}, respectively, showing
the compatibility between the calculation of the velocity centroid map from the PPV cube and 
directly from the simulation velocity and density cubes.
The isotropy degree of ``PPV$^2$ centroids'', panels  {\it(j)--(l)}, are also comparable with panels {\it(d)--(f)} and  {\it(g)--(i)}
although there are notable differences particularly with the application of a density gradient and white noise (dashed and dotted lines, respectively)
due to large scale fluctuations or small scale fluctuations, respectively. As expected, the X LOS (left column) 
shows high isotropy degree while the Y and Z LOS (center and right column) show low isotropy degree as they are perpendicular to the mean magnetic field.

We also investigate the application of velocity centroid structure function anisotropy technique as outlined above to
molecular emissions lines arising from the $^{13}$CO J2-1 transition.  
We apply the post-processing radiative transfer algorithm from Ossenkopf (2002) to our
MHD simulations.
We refer the reader to Ossenkopf (2002) and Burkhart et al. (2013 b, 2013c) 
for a detailed description of the radiative transfer algorithm.
We must scale the simulations to physical cloud parameters
and choose a similar initial set up to that of Burkhart et al. (2013 b, 2013c): a
cloud size of 5 pc, an average density of 275 cm$^{-3}$, the LOS perpendicular to the mean magnetic field, and a gas
temperature of 10K. The cube is observed at a distance of 450 pc with a beam FWHM of 18'' and a velocity resolution of 0.5km/s and the CO
abundance is $x_{co}=1.5x10^{-6}$.
The average optical depth of these simulations is slightly greater than unity.  

\begin{figure}
   \centering
   \includegraphics[scale=0.5]{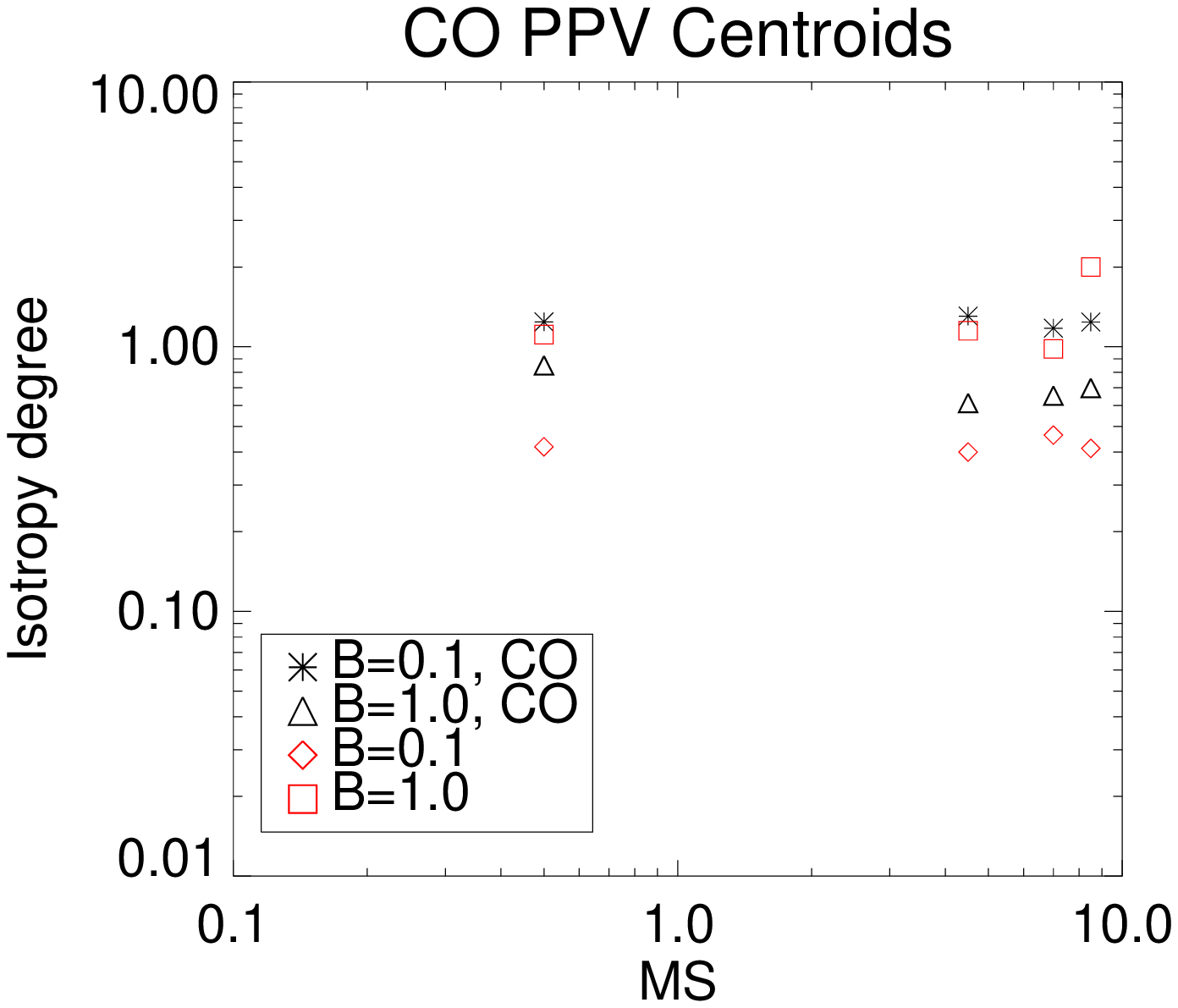}
   \caption{Degree of anisotropy in simulations with post-processing to include radiative transfer effects from $^{13}$CO emission (black triangles and stars)
as compared with simulations without radiative transfer (red squares and diamonds).
In both cases, the structure function of the velocity centroid is
averaged over scales from 10 grid points to 100 grid points.  
The X-axis corresponds to the sonic Mach number, and the Alfv\'enic Mach number is indicated 
by the various symbols as shown in the label. }
   \label{co}
   \end{figure}

We show the average isotropy degree vs. sonic Mach number of the structure functions of the CO velocity centroids in Figure \ref{co}.  
From our full parameter space represented in Table 1 we choose four different sonic Mach numbers ($\approx$ 8.5, 6.5, 3.5, 0.4) and two different initial Alfv\'enic number ($\approx $0.7 and 7.0) 
in order to cover the bulk of the parameter space that was shown in Figure 2. We also overplot the same synthetic velocity centroid isotropy degrees without radiative transfer effects

Figure \ref{co} shows very similar behavior and good agreement with the models of Figure 2 (represented as red diamond and square symbols), which do not include radiative transfer or spatial smoothing.  
Sub-Alfv\'enic CO emission creates considerable anisotropy in the isocontours of the CO velocity centroid maps, although it is slightly closer to the isotropic
case then the fully optically thin emission.   Super-Alfv\'enic CO emission remains
 isotropic.  This effect is largely insensitive to the sonic Mach number.
This gives us confidence that the method could be applied to observational CO emission cubes with success. We will test the effects of varying opacity in future works.

\subsection{Anisotropy for different LOS orientation angles}
\label{sublos}

\begin{figure*}
   \centering

   \includegraphics[width=5.2cm]{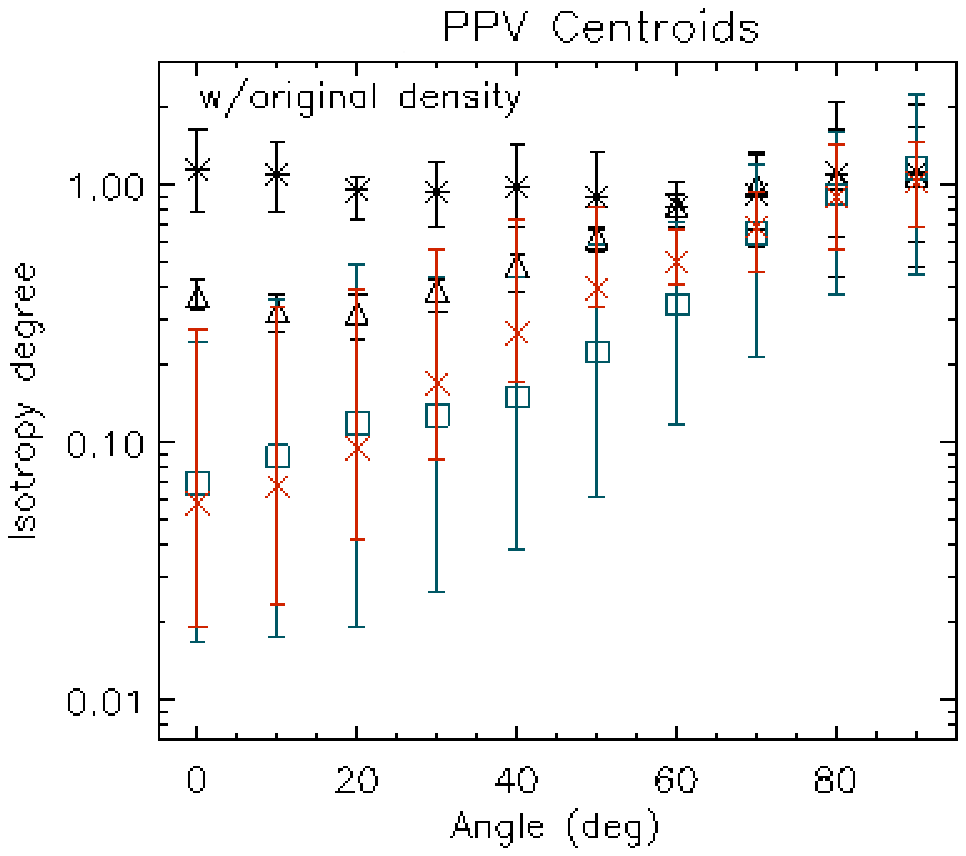}
   \includegraphics[width=5.2cm]{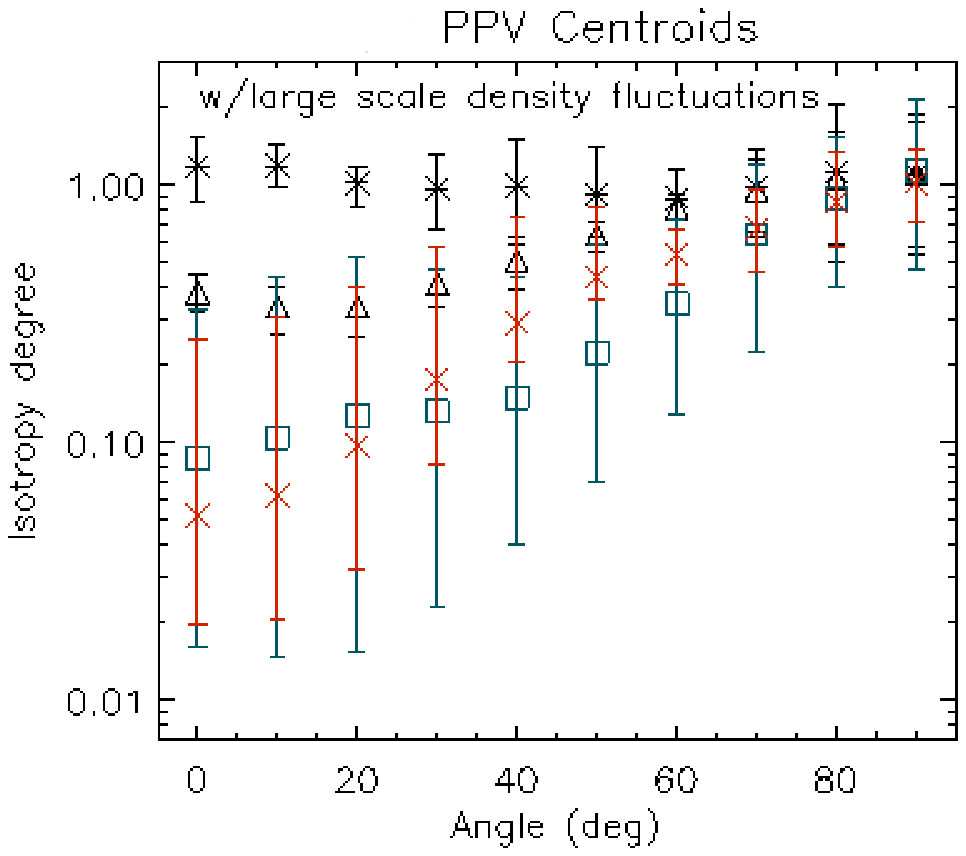}
   \includegraphics[width=5.2cm]{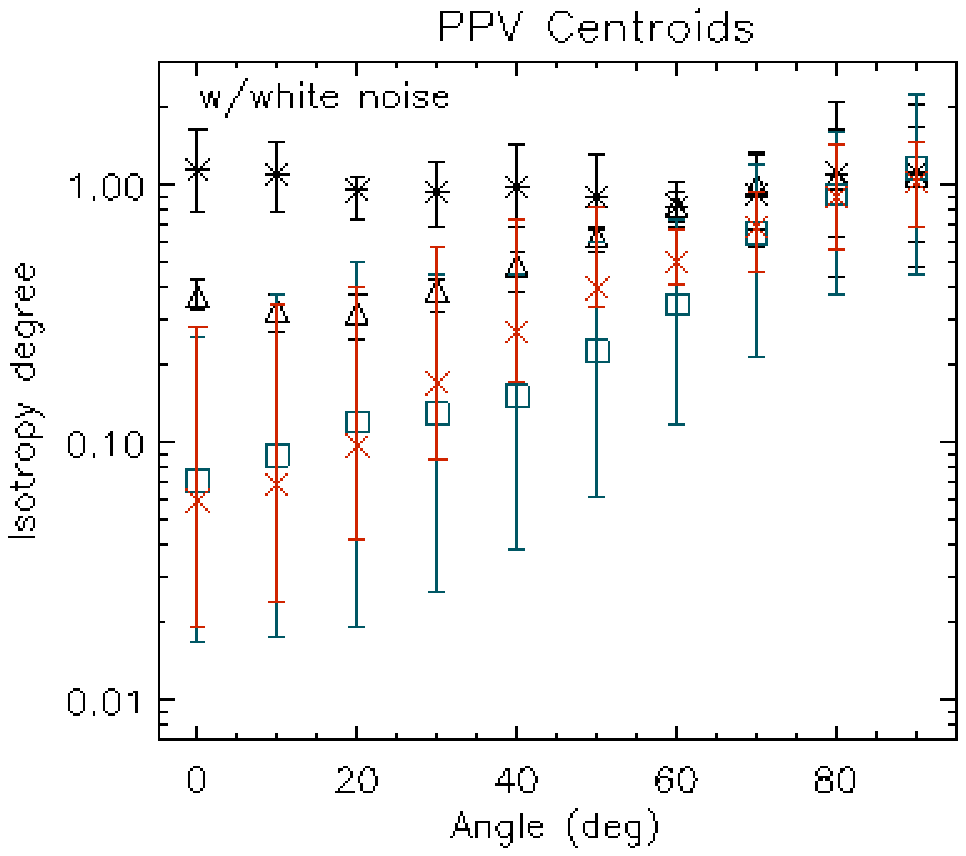}
  \includegraphics[width=5.2cm]{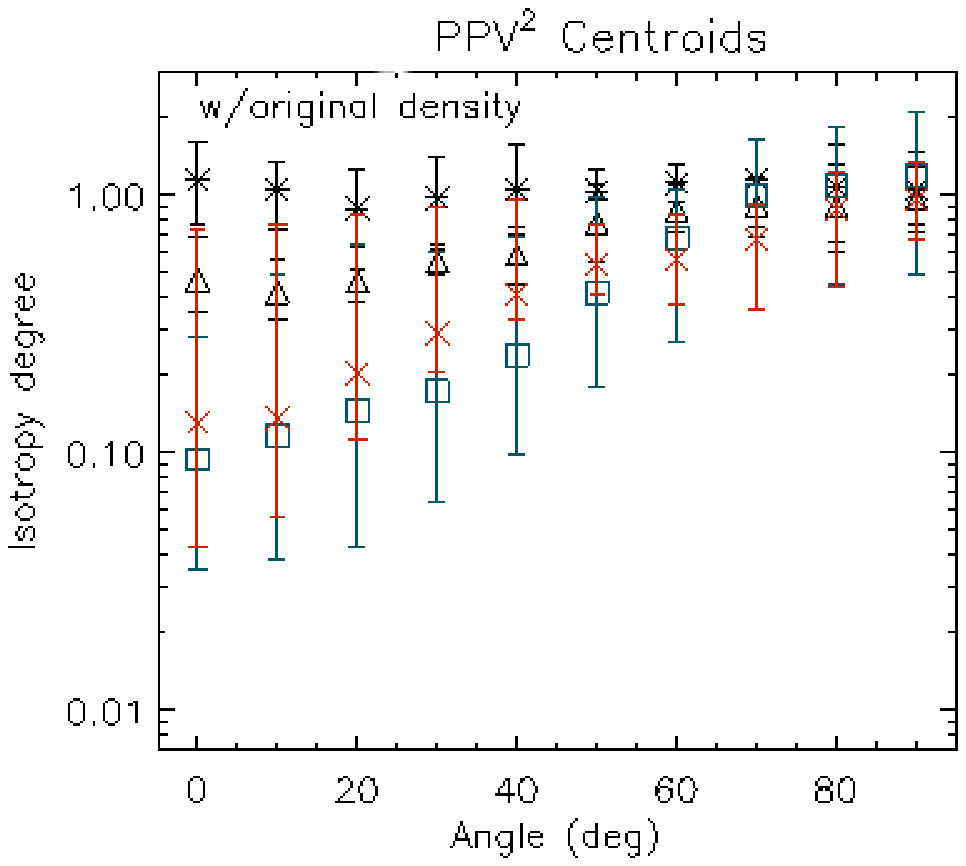}
  \includegraphics[width=5.2cm]{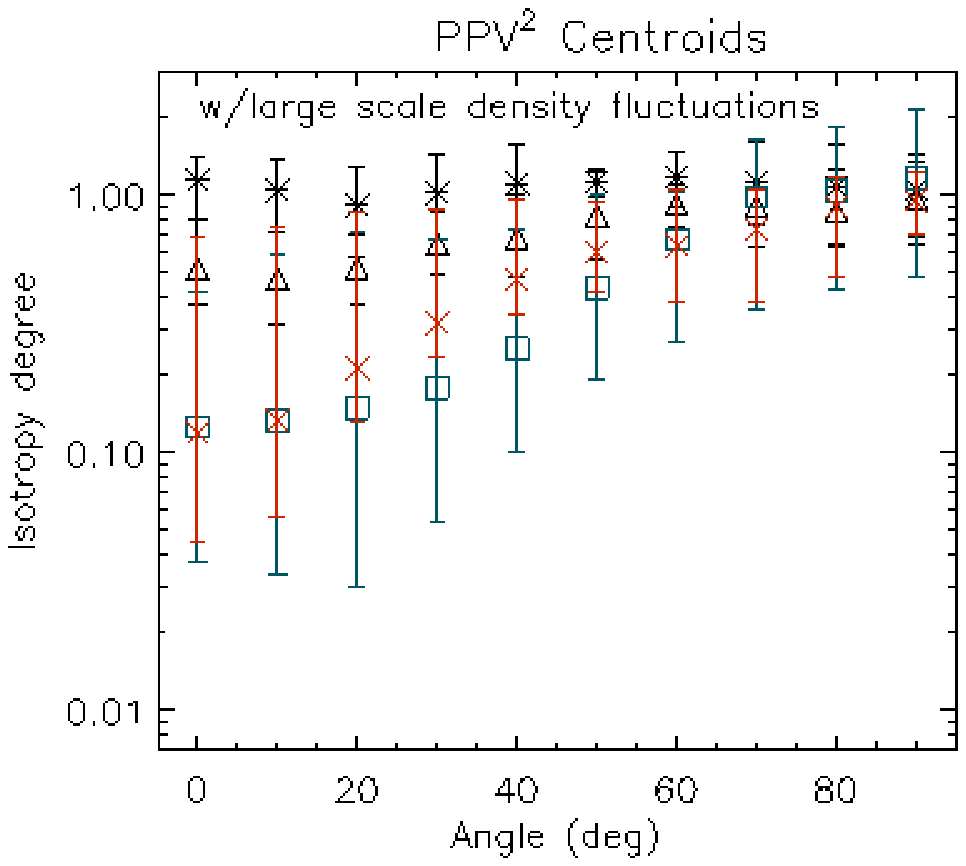}
  \includegraphics[width=5.2cm]{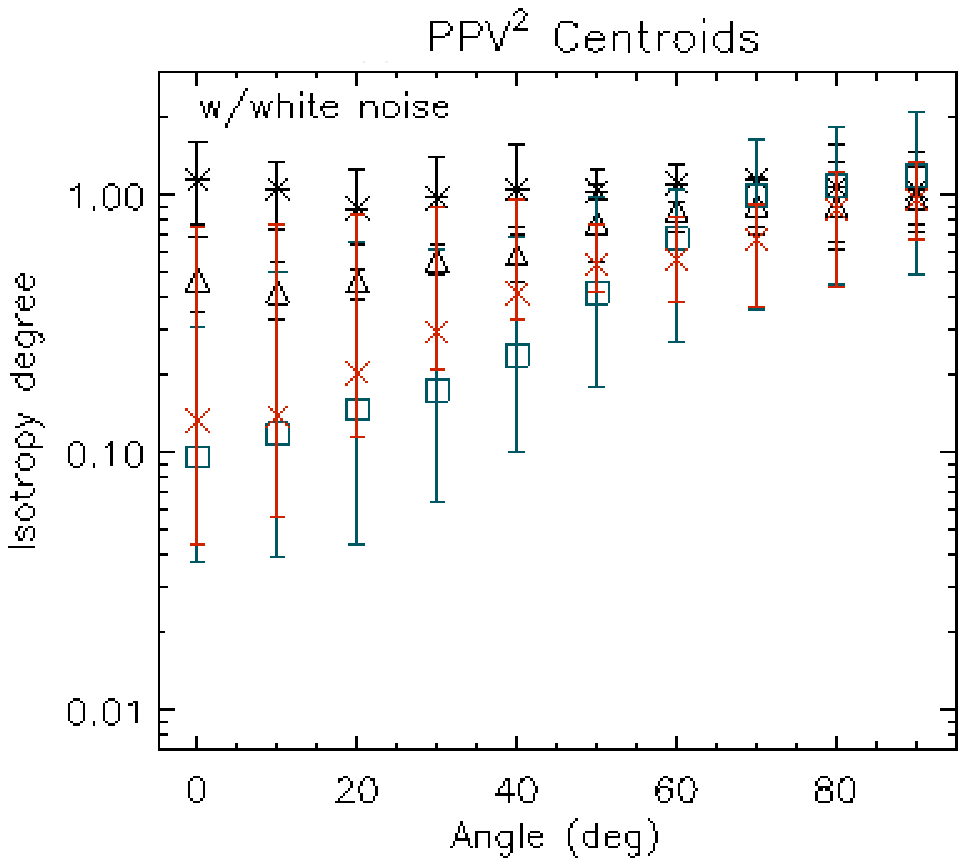}
   \caption{Same organization in Fig.~\ref{fig4}, but here the horizontal axis corresponds to the LOS orientation angle in respect of a perpendicular direction to the mean magnetic field.
The results are an average of two cases, where the LOS was rotated (i) around the $z$-axis (with angle = 0$^{\rm o}$ in the $y$-axis) and (ii) around the $y$-axis (with angle = 0$^{\rm o}$ in the $z$-axis).
For all these data points, $\langle{P_{{\rm gas},0}}\rangle$~$\sim$1.0 ($M_s$~$\sim$0.6; subsonic).
}
   \label{figang}
   \end{figure*}

\begin{figure*}
   \centering

   \includegraphics[width=5.2cm]{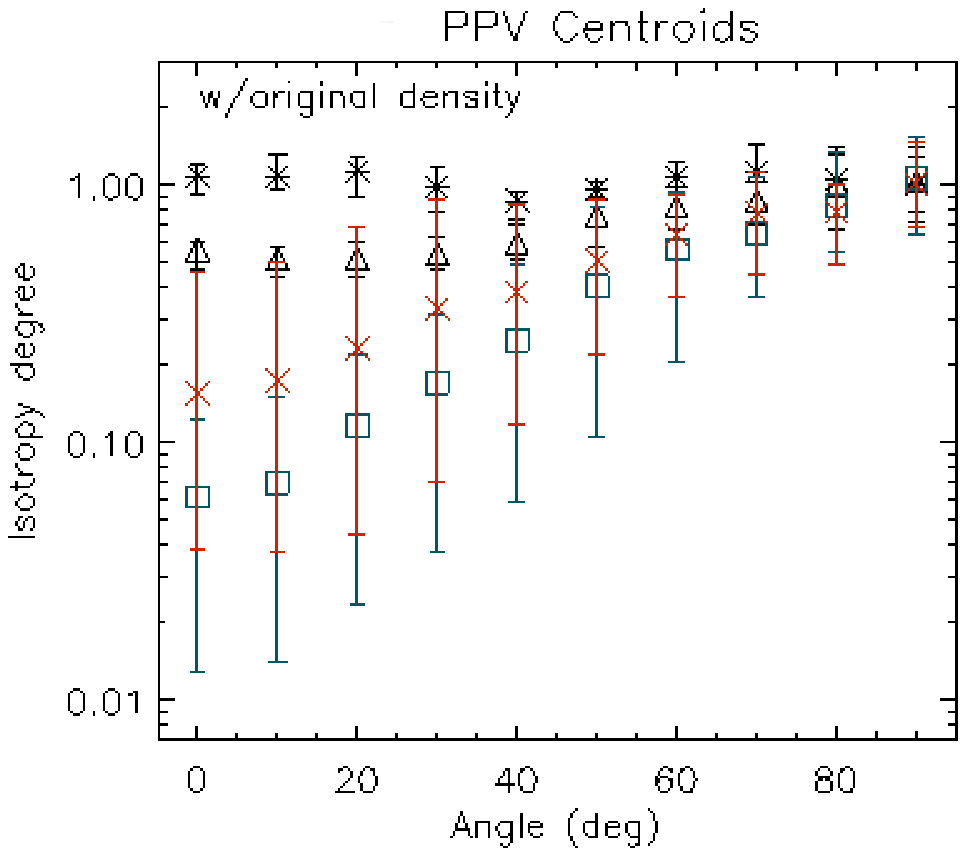} 
  \includegraphics[width=5.2cm]{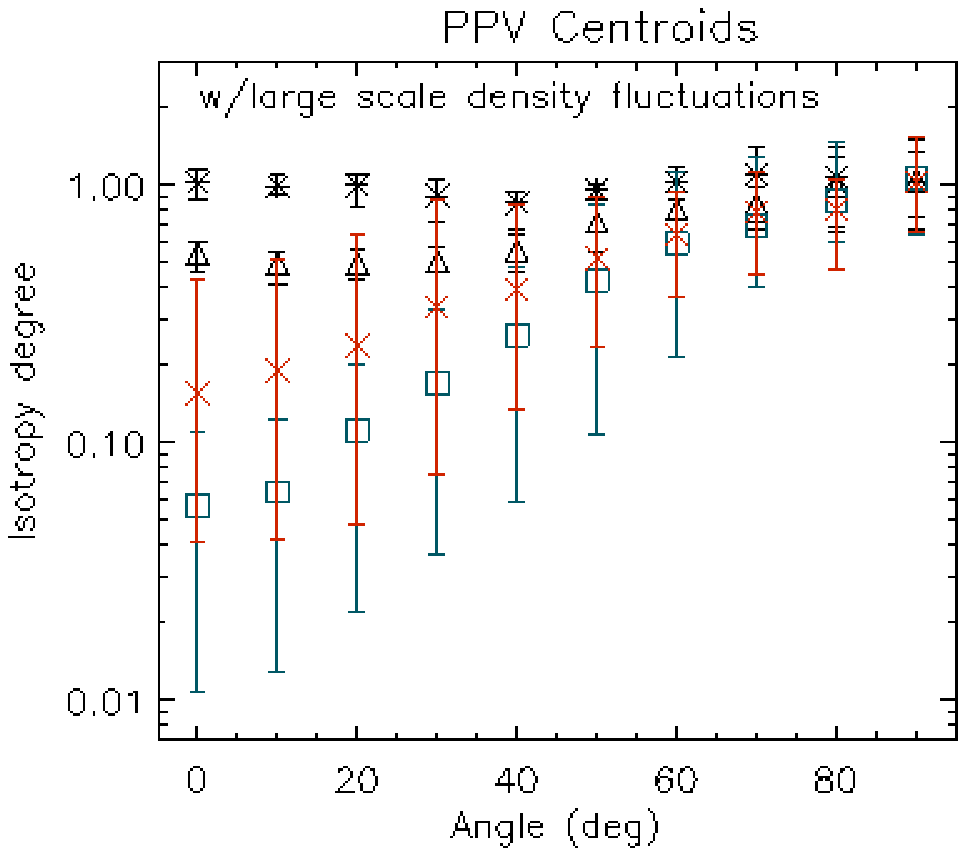}
   \includegraphics[width=5.2cm]{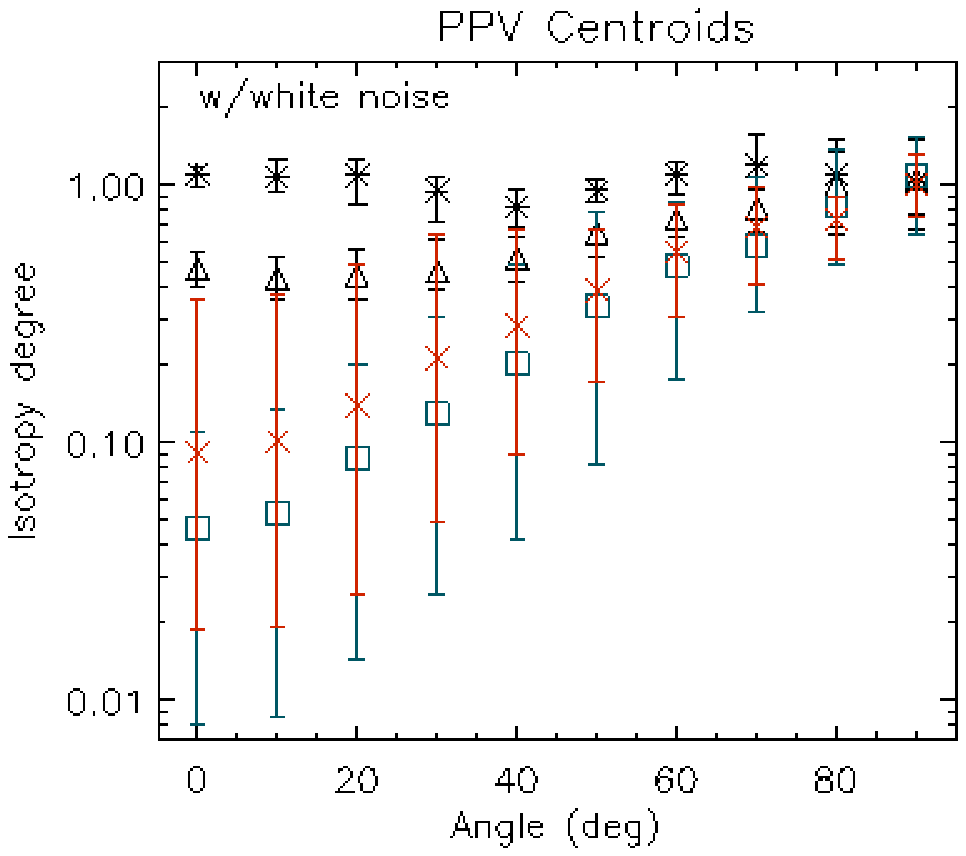}
   \includegraphics[width=5.2cm]{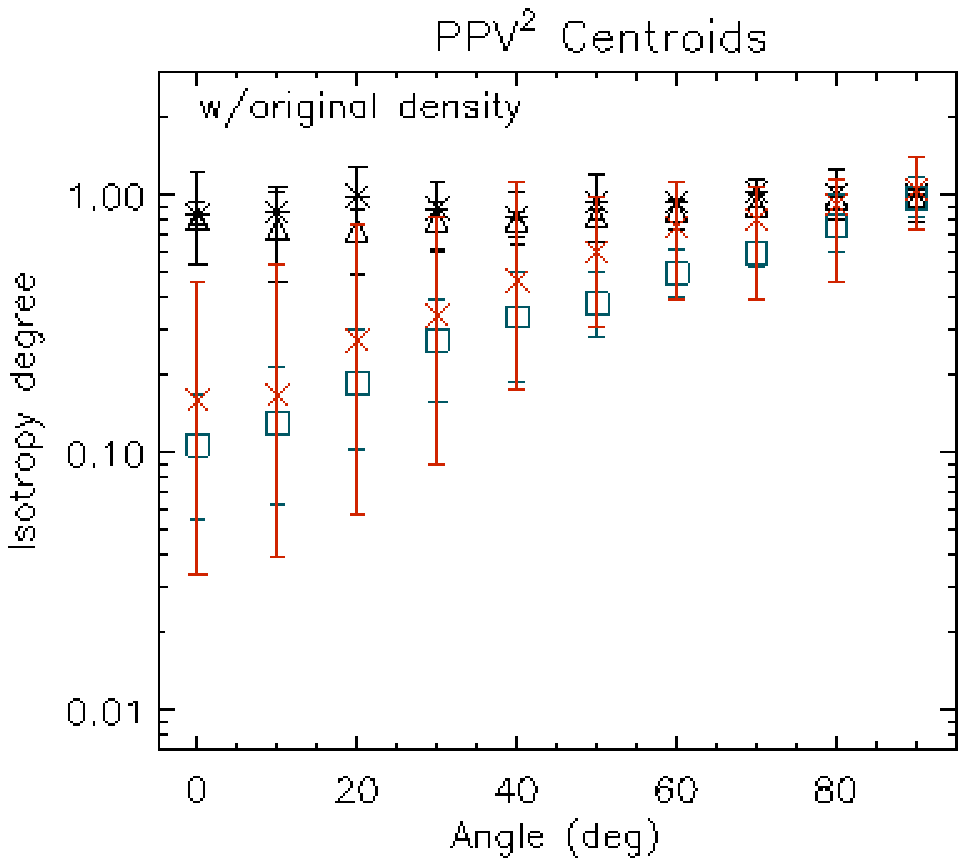}
   \includegraphics[width=5.2cm]{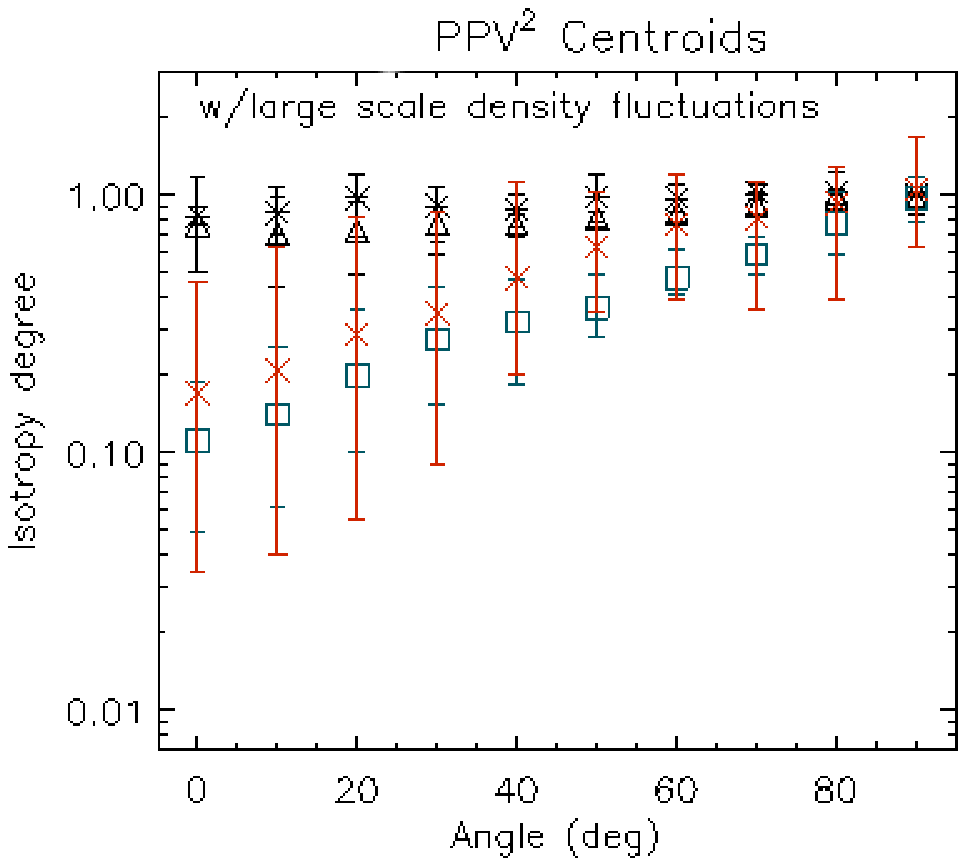}
   \includegraphics[width=5.2cm]{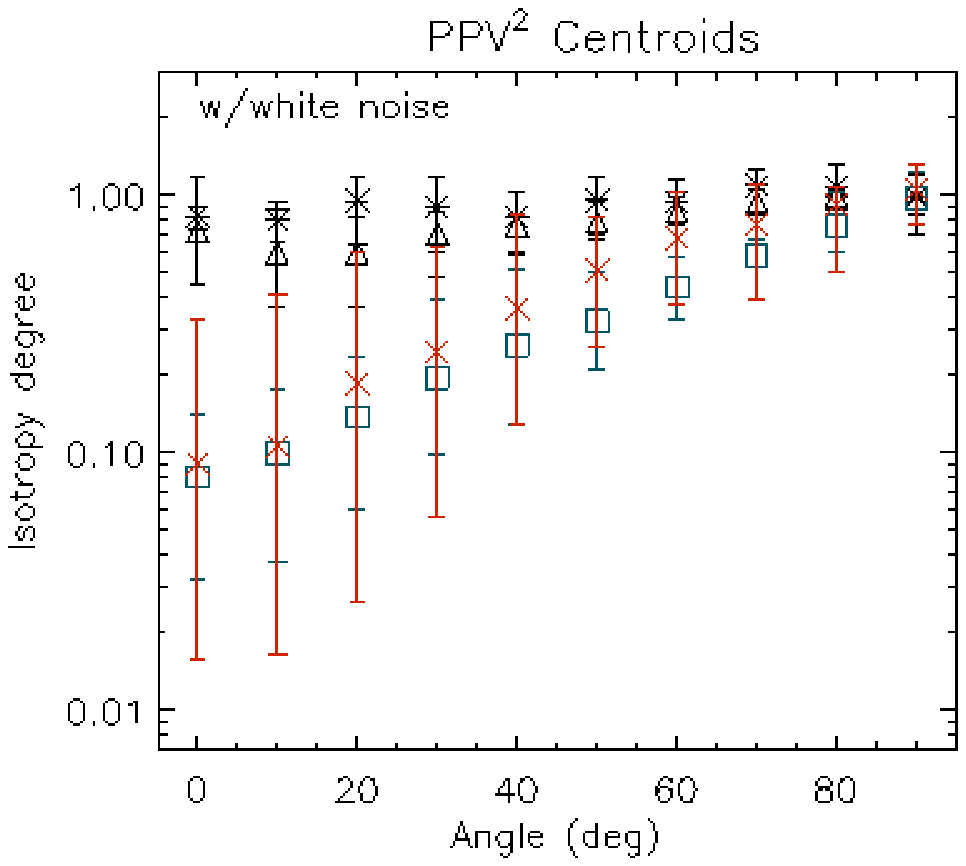}
   \caption{Same as in Fig.~\ref{figang}, but here $\langle{P_{{\rm gas},0}}\rangle$~$\sim$0.01 ($M_s$~$\sim$8.0; supersonic) for all the data points.}
   \label{figang2}
   \end{figure*}

EL11 only considered the LOS either parallel or perpendicular to the mean magnetic field.  However, the ISM has a range of LOS orientations relative to the local or global mean field. 
We repeat the analysis now including rotation of the LOS around an azimuthal origin (0$^{\rm o}$) and computed the parameter averages.
The LOS was rotated (i) around the $z$-axis (with the azimuthal origin found in the $y$-axis) and (ii) around the $y$-axis (with the azimuthal origin found in the $z$-axis).
Angle zero thus represents the LOS perpendicular to the mean magnetic field and angle 90 is along the x-axis (LOS parallel to the mean magnetic field).
These results are shown in Figs.~\ref{figang} and~\ref{figang2},
where in the upper row we considered the velocity centroids as those by EL11 (``ideal centroids''),
in the middle row we obtained the centroids from PPV cubes
and in the bottom row we used the ``PPV$^2$ centroids''.
In Fig.~\ref{figang} the isotropy degree is plotted vs. the LOS orientation angle for different $M_A$ values
and for a fixed $M_s \sim 0.6$, ie. in the subsonic regime.
In Fig.~\ref{figang2} the plots are shown for a fixed $M_s \sim 8.0$ in a supersonic regime (only considering velocity centroids
as the three different centroid calculations yield similar plots).

\begin{figure}
   \centering
   \includegraphics[scale=0.35]{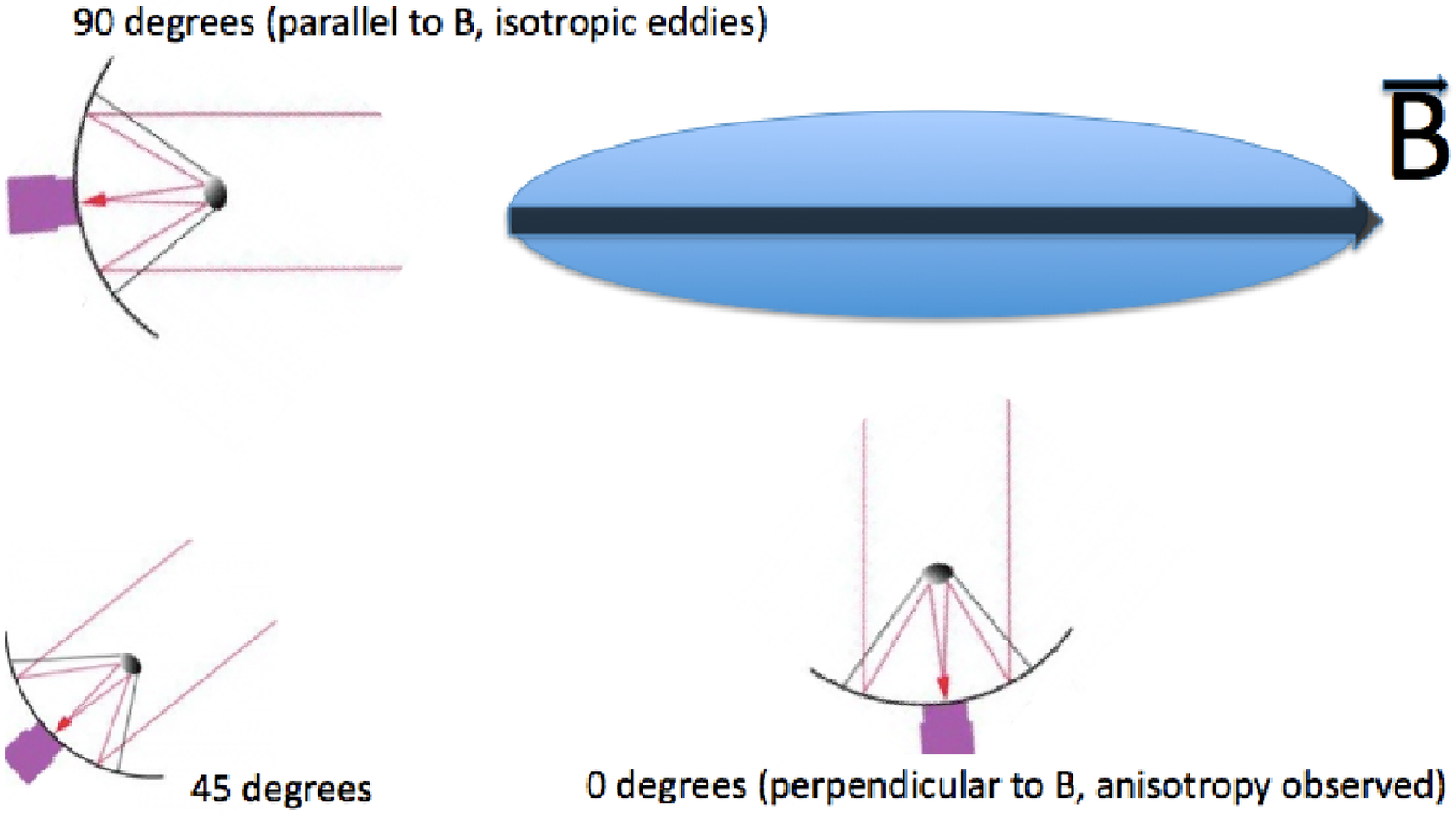}
   \caption{Cartoon illustration of the effect of the LOS on the observed anisotropy of an eddy which is elongated along the mean magnetic field line.
An observer looking perpendicular to the mean field will see anisotropy if the turbulence is trans- or sub-Alfv\'enic.  
An observer looking parallel to  will see an isotropic eddy regardless of field strength.   }
   \label{cartoon}
   \end{figure}

The observed velocity centroid anisotropy is greatest for sight-lines perpendicular
to the mean magnetic field (i.e. at zero degrees) regardless of sonic Mach number.
For super-Alfv\'enic turbulence, the turbulence remains isotropic regardless of observer angle
relative to the mean field.
For sub-Alfv\'enic turbulence, the observer LOS greatly alters the degree
of anisotropy observed in the velocity centroid map. In general, the method provides the possibility to distinguish between
sub-Alfv\'enic, super-Alfv\'enic and trans-Alfv\'enic turbulence from zero to forty degrees
relative to the axis perpendicular to the mean mean field.  Past this angle the turbulence
can begin to look isotropic despite the strength of the field.  At 90 degrees (i.e. parallel to the mean field)
the eddies in the centroid maps look isotropic regardless of magnetic field strength.  We illustrate
this in a cartoon in Figure ~\ref{cartoon}

\section{Discussion}
\label{secdisc}

The magnetic field in the ISM can be measured from different techniques,
as for example, Zeeman Splitting (e.g. molecular OH), star light polarization, and Faraday rotation (e.g. in HI).
However, these techniques often involve complex and difficult data reduction as well as 
large amounts of telescope time.
The method presented here allows one to estimate the plane-of-sky Alfv\'enic Mach number in the ISM from PPV velocity
centroid data, which are provided by several publically available surveys across multiple tracers (e.g. the COMPLETE survey, see Ridge et al. 2006). In this work 
we extended the original analysis of EL11 to include thermal broadening, a varying LOS relative to the mean
magnetic field, and investigated the anisotropy in the second moment maps.  We also tested our method on synthetic CO maps and found
that the effect is preserved even when telescope beam smoothing and radiative transfer effects are included.

The statistical measurement of the Alfv\'enic Mach number by observing anisotropy in velocity centroid maps is not without its limitations.  
For example, in this work we demonstrated that we can only detect a lower limit of anisotropy due to projection effects.  
That is, the mean isotropy degree increases and trend to $1.0$ when the LOS changes
from a perpendicular direction (angle = 0$^{\rm o}$) to the mean magnetic field
to a parallel direction (angle = 90$^{\rm o}$). However, dust polarization can provide complimentary plane-of-sky magnetic field directions to the method presented here.
The velocity centroid method presented in this paper should be compared with estimates of the Alfv\'enic Mach number from the Chandrasekhar-Fermi
technique when possible. The LOS field can not be obtained using anisotropy of the velocity centroids and other methods, such as
rotation measure or Zeeman splitting, should be employed.
 In general, we advocate that statistical techniques for studies of turbulence, including studying the Alfv\'enic Mach number,
should be done with with multiple techniques in mind.  For the estimation of the Alfv\'enic nature of the ISM this includes techniques
such as the bispectrum, PCA and phase coherence (Heyer et al. 2008; Burkhart et al. 2009; Burkhart \& Lazarian 2014, in prep.).

\section{Conclusions}
\label{seccon}

In this work we explored the anisotropy present in velocity centroid maps caused by an external mean magnetic field.
We used a set of ideal MHD simulations to create synthetic PPV maps with a wide range of sonic and Alfv\'enic Mach numbers and applied
the structure function to the velocity centroid maps.
We tested the use of our method for observations by including several effects present in the real data such as thermal 
line broadening, cloud boundaries, noise, and radiative transfer effects from the $^{13}$CO J2-1 transition.
We found that none of these effects altered greatly the anisotropy observed in the structure function contours.
However, the LOS relative to the mean magnetic field does alter the anisotropy observed.

We conclude that anisotropy in velocity centroids is a robust means of distinguishing the Alfv\'enic regime of a cloud
so long as the LOS relative to the global field is known.  
Since for a given LOS the anisotropy can only be equal or smaller than that observed when this is perpendicular to the mean magnetic field, 
the Alfv\'en Mach number obtained with our technique for any given LOS is an upper limit. 
In that sense, if one finds that the anisotropy is found to be consistent with sub-Alfv\'enic turbulence we can be confident that it is, 
however the opposite statement is not necessarily true.
Our method is complimentary to optical polarization 
measurements of the orientation (and strength via the Chandrasekhar-Fermi technique) to obtain the plane-of-sky magnetic field and 
methods to obtain the LOS field, 
such as Faraday rotation measurements.

\acknowledgments

B.B. acknowledges support from the Wisconsin Space Grant.
B.B. and A.L.  acknowledge support from the Center for Magnetic Self-Organization in Astrophysical and Laboratory Plasmas.
I. C. L. acknowledges a Post-Doctoral 
fellowship of the CNPq. 
J. R. de Medeiros acknowledges CNPq and FAPERN Brazilian agencies and by the INCT-INEspa\c{c}o Brazilian institute. 
A.E. acknowledges support from CONACYT grant 167611, and UNAM DGAPA grant IG100214.
Authors thank Volker Ossenkopf for the use of the SimLine3D radiative transfer code.

\end{document}